\documentclass[logo,copyright]{msas}

\usepackage{csquotes}

\usepackage[backend=bibtex,style=alphabetic,minnames=2,maxnames=5,maxcitenames=2,maxbibnames=99,natbib=true]{biblatex}
\usepackage{amsmath}
\usepackage{tikz}
\usepackage{mathdots}
\usepackage{yhmath}
\usepackage{cancel}
\usepackage{color}
\usepackage{siunitx}
\usepackage{array}
\usepackage{multirow}
\usepackage{amssymb}
\usepackage{gensymb}
\usepackage{tabularx}
\usepackage{nicematrix}
\usepackage{enumitem}
\usepackage{extarrows}
\usepackage{caption}
\usepackage{subcaption}
\usepackage{float}
\usepackage{booktabs}
\usepackage{cuted}
\usepackage{capt-of}
\usepackage[capitalise]{cleveref}
\usetikzlibrary{fadings}
\usetikzlibrary{patterns}
\usetikzlibrary{shadows.blur}
\usetikzlibrary{shapes}

\definecolor{high}{HTML}{ef3b2c}  %
\definecolor{low}{HTML}{fff7bc}  %

\crefname{appsec}{Appendix Section}{Appendix Sections}
\crefname{appfig}{Appendix Figure}{Appendix Figures}
\crefname{apptab}{Appendix Table}{Appendix Tables}
\crefname{appeq}{Appendix Equation}{Appendix Equations}

\definecolor{darkpastelgreen}{rgb}{0.01, 0.75, 0.24}

\title{Scaling transformer neural networks for skillful and reliable medium-range weather forecasting}

\author[1]{Tung Nguyen}
\author[1,2]{Rohan Shah}
\author[1]{Hritik Bansal}
\author[3]{Troy Arcomano}
\author[3,4]{Romit Maulik}
\author[3]{Veerabhadra Kotamarthi}
\author[3]{Ian Foster}
\author[3]{Sandeep Madireddy}
\author[1]{Aditya Grover}

\affil[1]{UCLA}
\affil[2]{CMU}
\affil[3]{Argonne National Laboratory}
\affil[4]{Penn State University}

\newcommand{\name}[1]{Stormer}

\correspondingauthor{tungnd@cs.ucla.edu, adityag@cs.ucla.edu}

\begin{abstract}
Weather forecasting is a fundamental problem for anticipating and mitigating the impacts of climate change. Recently, data-driven approaches for weather forecasting based on deep learning have shown great promise, achieving accuracies that are competitive with operational systems. However, those methods often employ complex, customized architectures without sufficient ablation analysis, making it difficult to understand what truly contributes to their success. Here we introduce \name{}, a simple transformer model that achieves state-of-the-art performance on weather forecasting with minimal changes to the standard transformer backbone. We identify the key components of \name{} through careful empirical analyses, including weather-specific embedding, randomized dynamics forecast, and pressure-weighted loss. At the core of \name{} is a randomized forecasting objective that trains the model to forecast the weather dynamics over varying time intervals. During inference, this allows us to produce multiple forecasts for a target lead time and combine them to obtain better forecast accuracy. On WeatherBench~2, \name{} performs competitively at short to medium-range forecasts and outperforms current methods beyond 7 days, while requiring orders-of-magnitude less training data and compute. Additionally, we demonstrate \name{}'s favorable scaling properties, showing consistent improvements with increases in model size and training tokens. 
Code and checkpoints are available at \url{https://github.com/tung-nd/stormer}.
\end{abstract}

\begin{document}
\maketitle

\begin{figure}[ht]
\centering
    \includegraphics[width=1.0\textwidth]{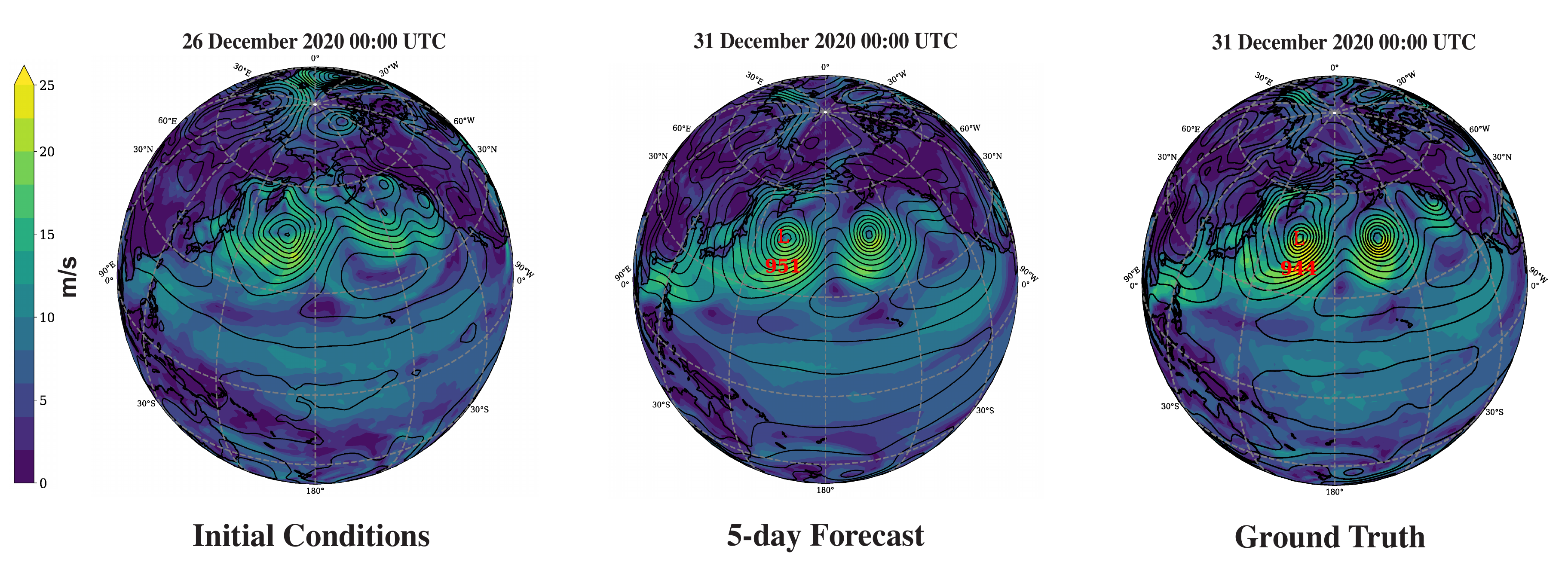}
    \caption{Illustration of an example $5$-day forecast of near-surface wind speed (color-fill) and mean sea level pressure (contours). On December 31, 2020, an extratropical cyclone impacted Alaska setting a new North Pacific low-pressure record. Here, we evaluate the ability of \name{} to predict this record-breaking event 5 days in advance. Using initial conditions from 0000 UTC, 26 December 2011, \name{} was able to successfully forecast both the location and strength of this extreme event with great accuracy.}
    \label{fig:Fig1}
\end{figure}

\section{Introduction}
\label{sec:intro}

Weather forecasting is a fundamental problem for science and society. With increasing concerns about climate change, accurate weather forecasting helps prepare and recover from the effects of natural disasters and extreme weather events, while serving as an important tool for researchers to better understand the atmosphere. Traditionally, atmospheric scientists have relied on numerical weather prediction (NWP) models~\citep{bauer2015quiet}. These models utilize systems of differential equations describing fluid flow and thermodynamics, which can be integrated over time to obtain future forecasts~\citep{lynch2008origins,bauer2015quiet}. Despite their widespread use, NWP models suffer from many challenges, such as parameterization errors of important small-scale physical processes, including cloud physics and radiation~\citep{stensrud2009parameterization}. Numerical methods also incur high computation costs due to the complexity of integrating a large system of differential equations, especially when modeling at fine spatial and temporal resolutions. Furthermore, NWP forecast accuracy does not improve with more data, as the models rely on the expertise of climate scientists to refine equations, parameterizations, and algorithms~\citep{magnusson2013factors}. 

To address the above challenges of NWP models, there has been an increasing interest in data-driven approaches based on deep learning for weather forecasting~\citep{gmd-11-3999-2018,scher2018toward,weyn2019can}. The key idea involves training deep neural networks to predict future weather conditions using historical data, such as the ERA5 reanalysis dataset~\citep{hersbach2018era5,hersbach2020era5,rasp2020weatherbench,rasp2023weatherbench}. Once trained, these models can produce forecasts in a few seconds, as opposed to the hours required by typical NWP models.
Because of the similar spatial structure between weather data and natural images,
early works in this space attempted to adopt standard vision architectures such as ResNet~\citep{rasp2021data,clare2021combining} and UNet~\citep{weyn2020improving} for weather forecasting, but their performances lagged behind those of numerical models. 
However, significant improvements have been made in recent years due to better model architectures and training recipes, and increasing data and compute~\citep{keisler2022forecasting,pathak2022fourcastnet,nguyen2023climax,bi2023accurate,lam2022graphcast,chen2023fengwu,chen2023fuxi}.
Pangu-Weather~\citep{bi2023accurate}, a 3D Earth-Specific Transformer model trained on 0.25$^{\circ}$ data (721$\times$1440 
grids), was the first model to outperform operational IFS~\citep{wedi2015modelling}. Shortly after, GraphCast~\citep{lam2022graphcast} scaled up the graph neural network architecture proposed by~\citet{keisler2022forecasting} to 0.25$^\circ$ data and showed improvements over Pangu-Weather. 
Despite impressive forecast accuracy, existing methods often involve highly customized neural network architectures with minimal ablation studies, making it difficult to identify which components contribute to their success. For example, it is unclear what the benefits of 3D Earth-Specific Transformer over a standard Transformer are, and how critical the multi-mesh message-passing in GraphCast is to its performance.
A deeper understanding, and ideally a simplification, of these existing approaches is essential for future progress in the field. Furthermore, establishing a common framework would facilitate the development of foundation models for weather and climate that extend beyond weather forecasting~\citep{nguyen2023climax}.

In this paper, we show that a simple architecture with a proper training recipe can achieve state-of-the-art performance. We start with a standard vision transformer (ViT) architecture, and through extensive ablation studies, identify the three key components to the performance of the model: (1) a weather-specific embedding layer that transforms the input data to a sequence of tokens by modeling the interactions among atmospheric variables; (2) a randomized dynamics forecasting objective that trains the model to predict the weather dynamics at random intervals; and (3) a pressure-weighted loss that weights variables at different pressure levels in the loss function to approximate the density at each pressure level. During inference, our proposed randomized dynamics forecasting objective allows a single model to produce multiple forecasts for a specified lead time by using different combinations of the intervals for which the model was trained. For example, one can obtain a 3-day forecast by rolling out the 6-hour predictions 12 times or 12-hour predictions 6 times. Combining these forecasts leads to significant accuracy improvements, especially for long lead times. We evaluate our proposed method, namely \textbf{S}calable \textbf{t}ransf\textbf{orm}ers for weath\textbf{er} forecasting (\name{}), on WeatherBench~2~\citep{rasp2023weatherbench}, a widely used benchmark for data-driven weather forecasting. \name{} achieves competitive forecast accuracy of key atmospheric variables for 1--7 days and outperforms the state-of-the-art beyond 7 days. Notably, \name{} achieves this performance by training on more than 5$\times$ lower-resolution data and orders-of-magnitude fewer GPU hours compared to the baselines.
Finally, our scaling analysis shows that the performance of \name{} improves consistently with increases in model capacity and data size, demonstrating the potential for further improvements.
\section{Background and Preliminaries}
\label{sec:problem_statement}
\begin{figure}[t]
    \centering
    \includegraphics[width=0.5\linewidth]{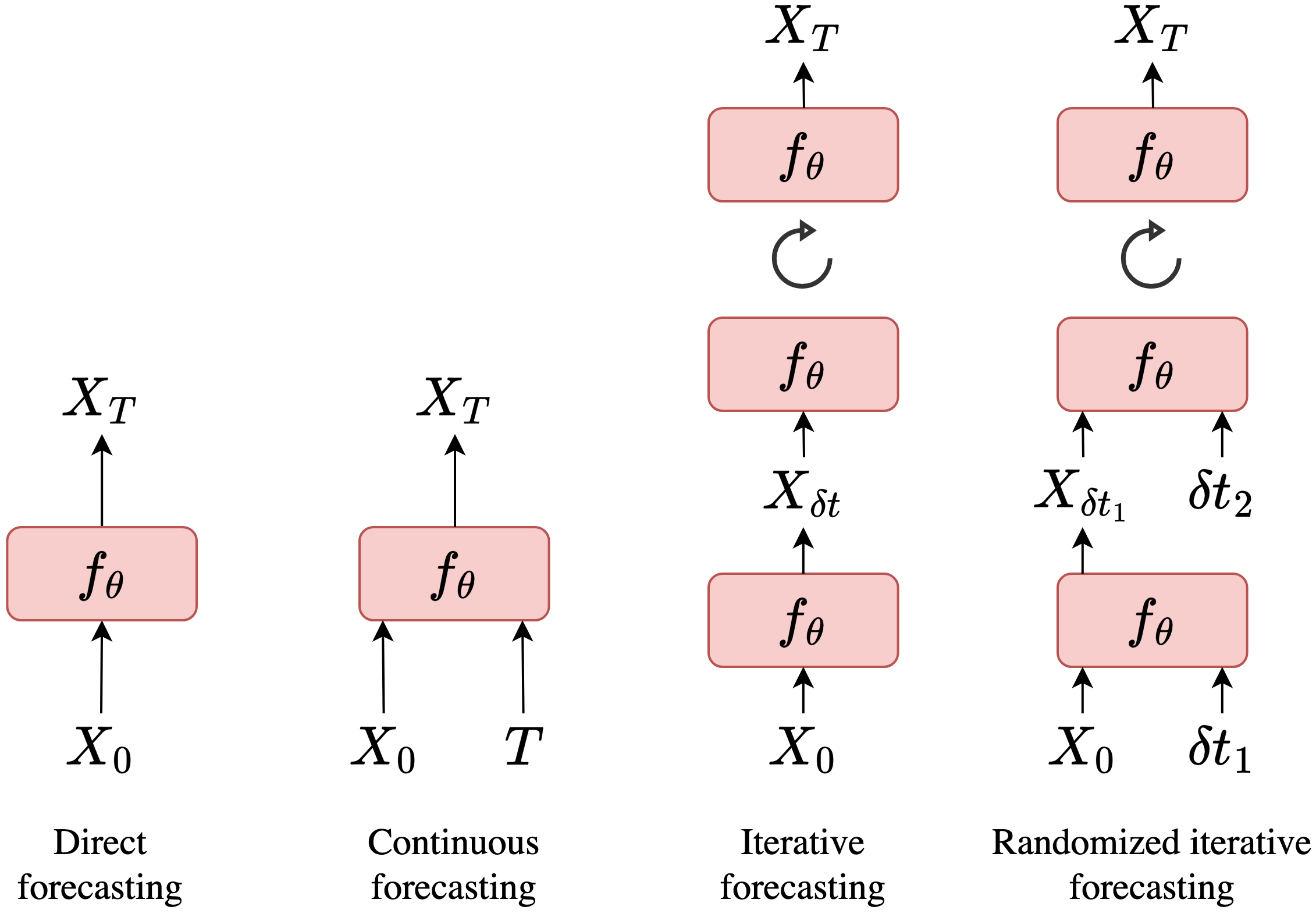}
    \caption{Different approaches to weather forecasting. 
    Direct and continuous methods output forecasts directly, but continuous forecasting is adaptable to various lead times by conditioning on \(T\). Iterative forecasting generates forecasts at small intervals \(\delta t\), which are rolled out for the final forecast. Our proposed randomized iterative forecasting combines continuous and iterative methods.}
    \label{fig:approach}
\end{figure}
\vspace{-5pt}
Given a dataset $\mathcal{D} = \{X_i\}_{i=1}^N$ of historical weather data, the task of global weather forecasting is to forecast future weather conditions $X_T \in \mathbb{R}^{V \times H \times W}$ given  initial conditions $X_0 \in \mathbb{R}^{V \times H \times W}$, in which $T$ is the target lead time, e.g., 7 days; $V$ is the number of input and output atmospheric variables, such as temperature and humidity;
and $H \times W$ is the spatial resolution of the data, which depends on how densely we grid the globe.  This formulation is similar to many image-to-image tasks in computer vision such as segmentation or video frame prediction. However, unlike the RGB channels in natural images, weather data can contain up to $100$s of channels. These channels represent actual physical variables that can be unbounded in values and follow complex laws governed by atmospheric physics. Therefore, the ability to model the spatial and temporal correlations between these variables is crucial to forecasting.

There are three major approaches to data-driven weather forecasting.  The first and simplest is \emph{direct forecasting}, which trains the model to directly output future weather $\widehat{X}_T = f_\theta(X_0)$ for each target lead time $T$. Most early works in the field adopt this approach~\citep{gmd-11-3999-2018,scher2018toward,weyn2019can,rasp2021data,clare2021combining,weyn2020improving}. Since the weather is a chaotic system, forecasting the future directly for large $T$ is challenging, which may explain the poor performances of these early models. Moreover, direct forecasting requires training one neural network for each lead time, which can be computationally expensive when the number of target lead times increases. To avoid the latter issue, \textit{continuous forecasting} uses $T$ as an additional input: $\widehat{X}_T = f_\theta(X_0, T)$, allowing a single model to produce forecasts at any target lead time after training. MetNet~\citep{sonderby2020metnet,espeholt2022deep,andrychowicz2023deep} employed the continuous approach for nowcasting at different lead times up to $24$ hours, WeatherBench~\citep{rasp2021data} considered continuous forecasting as one of the baselines, and ClimaX~\citep{nguyen2023climax} used this approach for pretraining.  However, since this approach still attempts to forecast future weather directly, it suffers from the same challenging problem of forecasting the chaotic weather in one step.
Finally, \textit{iterative forecasting} trains the model to produce forecasts at a small interval $\widehat{X}_{\delta t} = f_\theta(X_0)$, in which $\delta t$ is typically from 6 to 24 hours. To produce longer-horizon forecasts, we roll out the model by iteratively feeding its predictions back in as input. This is a common paradigm in both traditional NWP systems and the two state-of-the-art deep learning methods, Pangu-Weather and GraphCast. One drawback of this approach is error accumulation when the number of rollout steps increases, which can be mitigated by a multi-step loss function~\citep{keisler2022forecasting,lam2022graphcast,chen2023fengwu,chen2023fuxi}. In iterative forecasting, one can forecast either the weather conditions $X_{\delta t}$ or the weather dynamics $\Delta_{\delta t} = X_{\delta t} - X_0$, 
and $X_{\delta t}$ can be recovered by adding the predicted dynamics to the initial conditions. In this work, we adopt 
the latter approach,
which we refer to as \emph{iterative dynamics forecasting}. Figure~\ref{fig:approach} summarizes these different approaches.
\section{Methodology}
\label{sec:method}

We introduce \name{}, a skillful method for weather forecasting, and show that a simple architecture can achieve competitive forecast performances with a well-designed framework. We first present the overall training and inference procedure of \name{}, then describe the model architecture we use in practice. Section~\ref{sec:ablation} empirically demonstrates the importance of each component of \name{}.

\subsection{Training} \label{sec:training}
\begin{figure*}
  \centering
  \begin{subfigure}{0.3\linewidth}
    \includegraphics[width=1.0\linewidth]{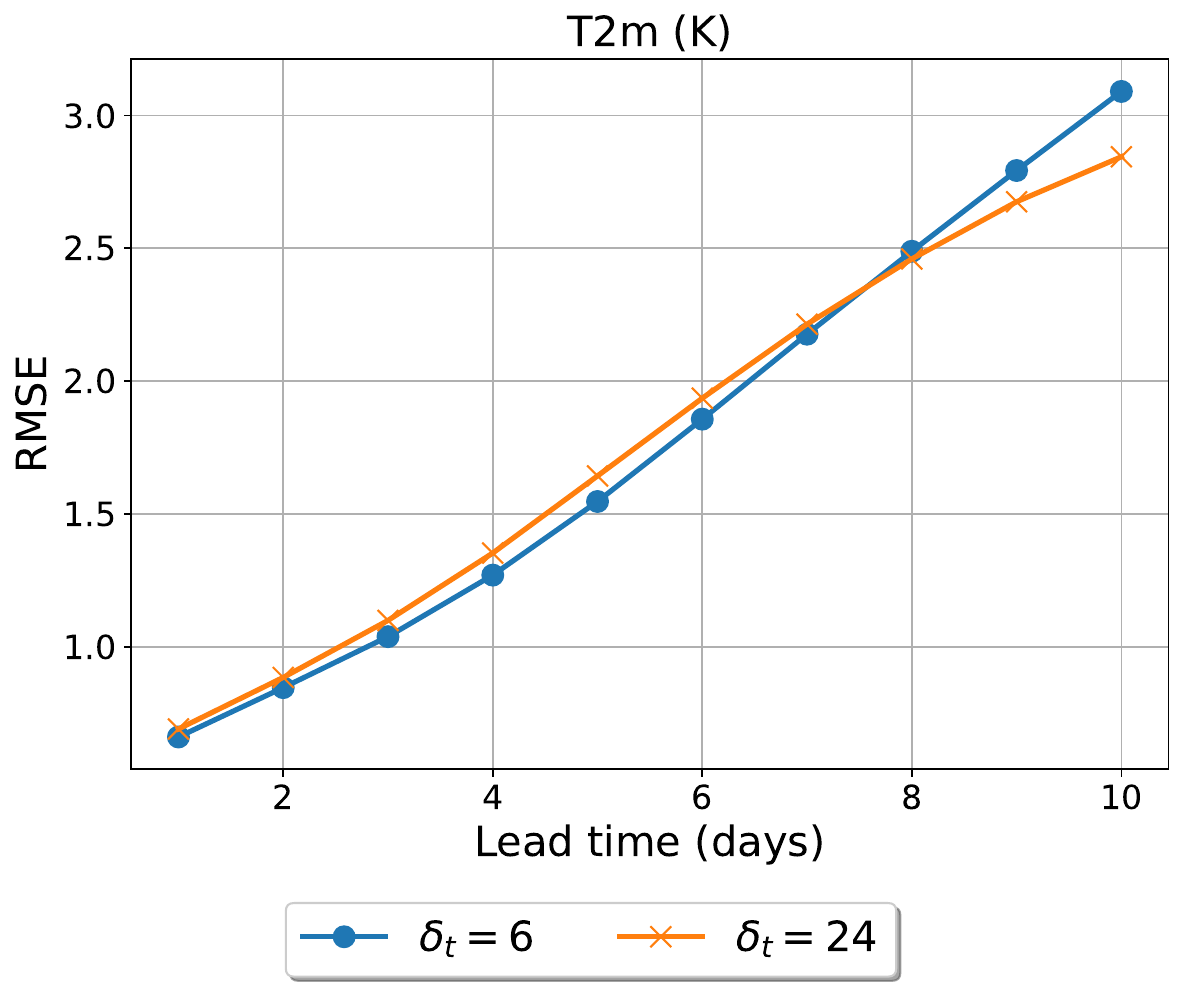}
    \caption{Time interval comparison.}
    \label{fig:ablation_interval}
  \end{subfigure}
  \hfill
  \begin{subfigure}{0.3\linewidth}
    \includegraphics[width=1.0\linewidth]{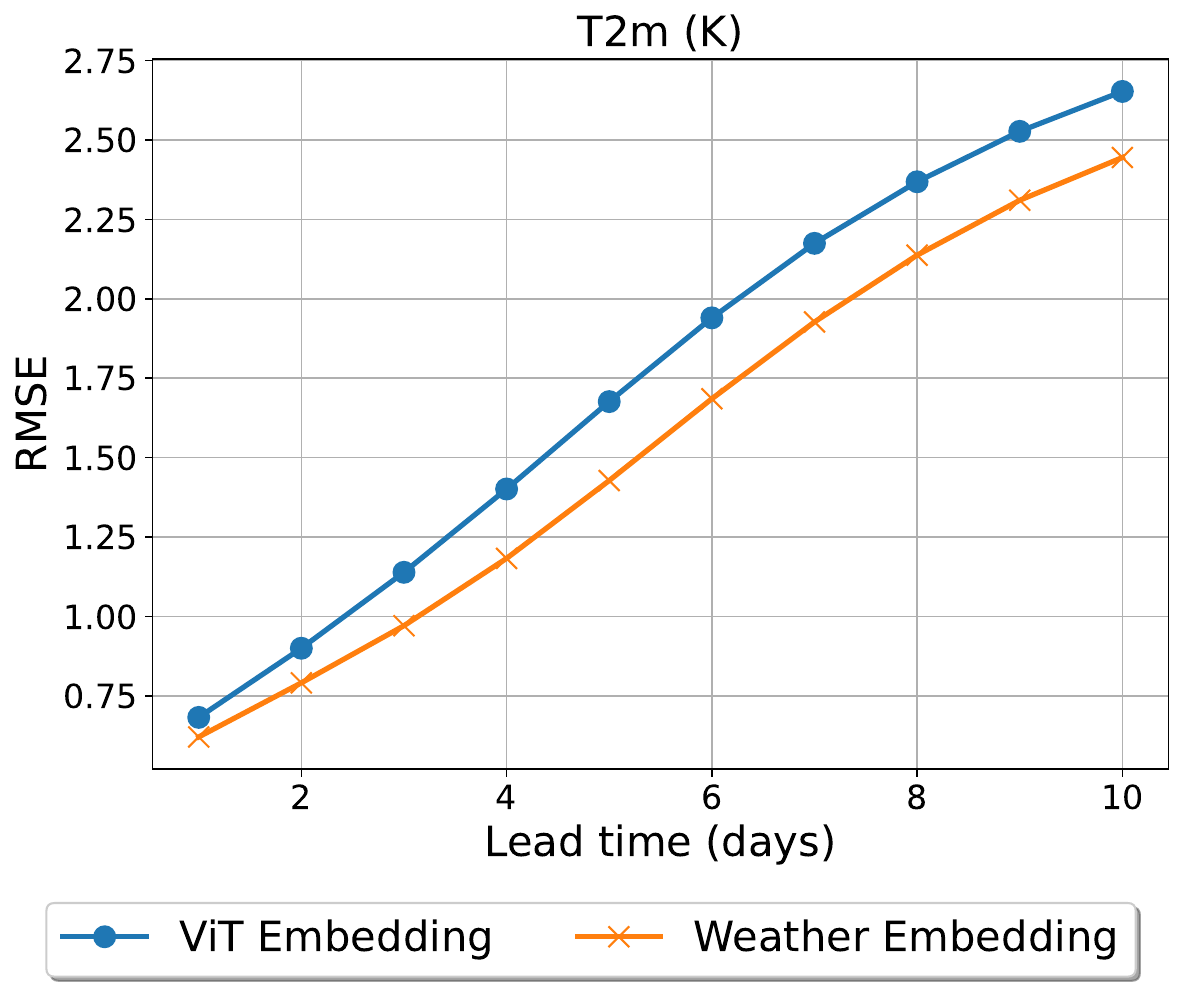}
    \caption{Patch embedding comparison.}
    \label{fig:compare_embeding}
  \end{subfigure}
  \hfill
  \begin{subfigure}{0.3\linewidth}
    \includegraphics[width=1.0\linewidth]{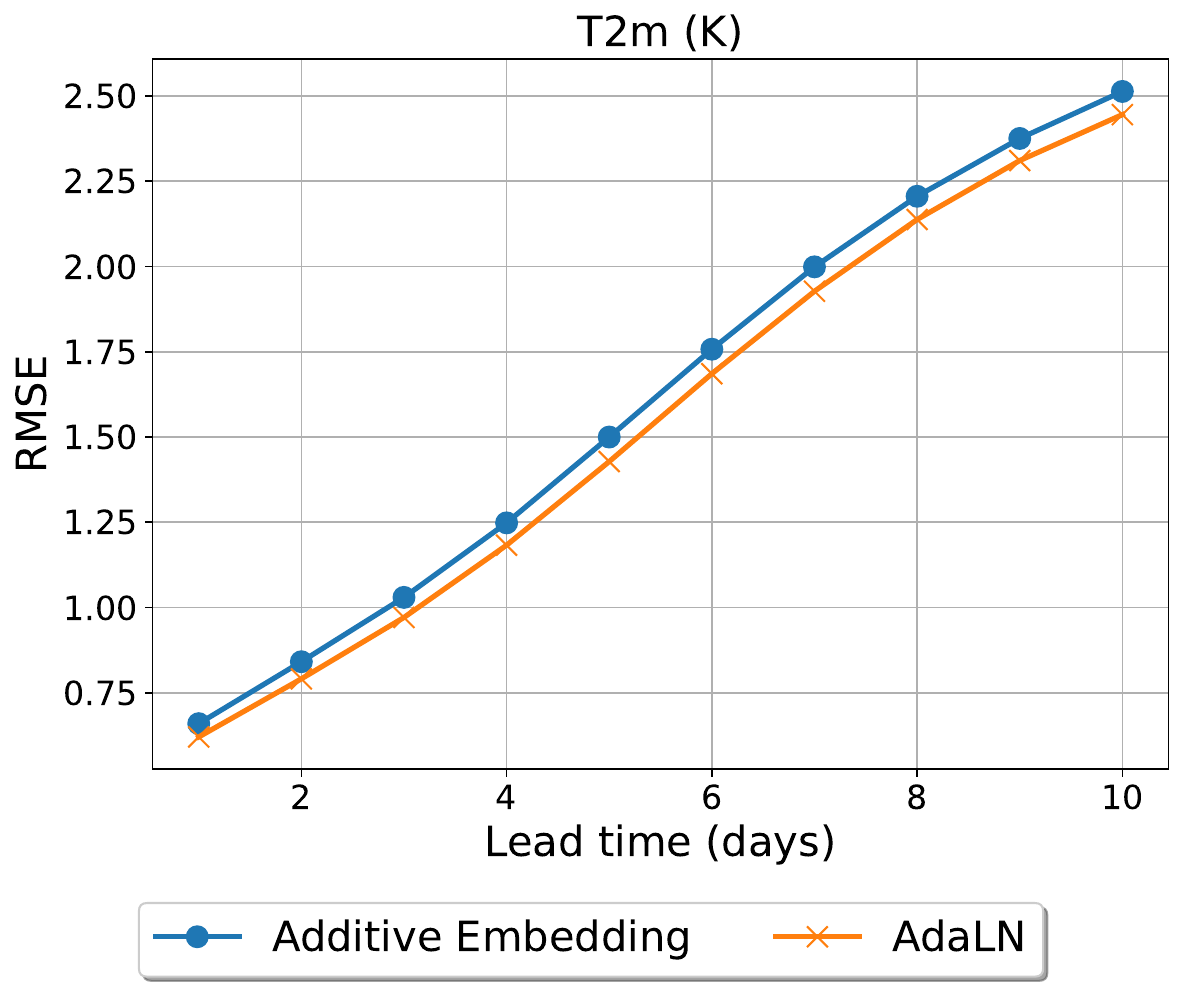}
    \caption{Time embedding comparison.}
    \label{fig:adaln}
  \end{subfigure}
  \caption{Preliminary results on forecasting surface temperature that led to the design choices of \name{}: (a) Different intervals are better at different lead times, (b) Weather-specific embedding is superior to standard ViT embedding, and (c) Adaptive layer norm outperforms additive embedding.}
  \label{fig:design_choices}
\end{figure*}

We adopt the iterative approach for \name{}, and train the model to forecast the weather dynamics $\Delta_{\delta t} = X_{\delta t} - X_0$, which is the difference between two consecutive weather conditions, $X_0$ and $X_{\delta t}$, across the time interval $\delta t$. A common practice in previous works~\citep{keisler2022forecasting,lam2022graphcast} is to use a small fixed value of $\delta t$ such as 6 hours. However, as we show in Figure~\ref{fig:ablation_interval}, while small intervals tend to work well for short lead times, larger intervals excel at longer lead times (beyond 7 days) due to less error accumulation.
Therefore, having a model that can produce forecasts at different intervals and combine them in an effective manner has the potential to improve the performance of single-interval models. This motivates our \emph{randomized dynamics forecasting objective}, which trains \name{} to forecast the dynamics at random intervals $\delta t$ by conditioning on $\delta t$:
\begin{equation}
    \small
    \mathcal{L}(\theta) = \mathbb{E}_{\delta t \sim P(\delta t), (X_0, X_{\delta t}) \sim \mathcal{D}}\left[||f_\theta(X_0, \delta t) - \Delta_{\delta t}||_2^2 \right],
\end{equation}
in which $P(\delta t)$ is the distribution of the random interval. 
In our experiments, unless otherwise specified, $P(\delta t)$ is a uniform distribution over three values $\delta t \sim \mathcal{U}\{\textrm{6}, \textrm{12}, \textrm{24}\}$. These three time intervals play an important role in atmospheric dynamics. The 6 and 12-hour values help to encourage the model to learn and resolve the diurnal cycle (day-night cycle), one of the most important oscillations in the atmosphere driving short-term dynamics (e.g., temperature over the course of a day). The 24-hour value filters the effects of the diurnal cycle and allows the model to learn longer, synoptic-scale dynamics which are particularly important for medium-range weather forecasting \citep{Holton2004}.

From a practical standpoint, this randomized objective provides two benefits. First, randomizing \(\delta t\) enlarges the training data, serving as data augmentation. Second, it allows a single trained model to generate various forecasts for a specified lead time $T$ by creating different combinations of intervals $\delta t$ that sum to $T$. For example, to forecast 7 days ahead, one could use 12-hour forecasts 14 times or 24-hour forecasts 7 times. Our experiments show that combining these forecasts is crucial for achieving good accuracy, especially for longer lead times. While both our approach and the continuous approach use the time interval as an additional input, we perform iterative forecasting instead of direct forecasting. This avoids the challenge of directly modeling chaotic weather and offers more flexibility for combining different intervals at test time.

\subsubsection{Pressure-weighted loss}
Due to the large number of variables being predicted, we use a physics-based weighting function to weigh variables near the surface higher. Since each variable lies on a specific pressure level, we can use pressure as a proxy for the density of the atmosphere at each level. This weighting allows the model to prioritize near-surface variables, which are important for weather forecasting and have the most societal impact. 
The final objective function that we use for training is:
\begin{equation}
    \small
    \mathcal{L}(\theta) = \mathbb{E}\left[ \frac{1}{VHW} \sum_{v=1}^V \sum_{i=1}^H \sum_{j=1}^W w(v) L(i) (\widehat{\Delta}_{\delta t}^{vij} - \Delta_{\delta t}^{vij})^2 \right]. \label{eq:obj}
\end{equation}
The expectation is over $\delta t, X_0$, and $X_{\delta t}$ which we omit for notational simplicity. 
In this equation, $w(v)$ is the weight of variable $v$, and $L(i)$ is the latitude-weighting factor commonly used in previous works to account for the non-uniformity when we grid the spherical globe~\citep{rasp2020weatherbench,keisler2022forecasting,pathak2022fourcastnet,nguyen2023climax,bi2023accurate,lam2022graphcast}. The pressure-weighted loss was first introduced by GraphCast~\citep{lam2022graphcast}, and we show that it also helps with a different architecture.

\subsubsection{Multi-step finetuning} \label{sec:multi_step}
To produce forecasts at a lead time beyond the training intervals, we roll out the model several times. Since the model's forecasts are fed back as input, the forecast error accumulates as we roll out more steps. To alleviate this issue, we finetune the model on a multi-step loss function. Specifically, for each gradient step, we roll out the model $K$ times, and average the objective~\eqref{eq:obj} over the $K$ steps:
\begin{equation}
    \small
    \mathcal{L}(\theta) = \mathbb{E}\left[ \frac{1}{KVHW} \sum_{k=1}^K \sum_{v=1}^V \sum_{i=1}^H \sum_{j=1}^W w(v) L(i) (\widehat{\Delta}_{k \delta t}^{vij} - \Delta_{k \delta t}^{vij})^2 \right]. \label{eq:finetune}
\end{equation}
In practice, we implement a three-phase training procedure for \name{}. In the first phase, we train the model to perform single-step forecasting, which is equivalent to optimizing the objective in~\eqref{eq:obj}. In the second and third phases, we finetune the trained model from the preceding phase with $K=$ 4 and $K=$ 8, respectively. We use the same sampled value of the interval $\delta t$ for all $K$ steps. We tried randomizing $\delta t$ at each rollout step, but found that doing so destabilized training as the loss value at each step is of different magnitudes, hurting the final performance of the model. Multi-step finetuning was used in FourCastNet~\citep{pathak2022fourcastnet} and also adopted in more recent works~\citep{keisler2022forecasting,lam2022graphcast}.

\subsection{Inference}
At test time, \name{} can produce forecasts at any time interval $\delta t$ used during training. Thus the model can generate multiple forecasts for a target lead time $T$ by creating different combinations of $\delta t$ that sum to $T$. We consider two inference strategies for generating forecasts:

\textbf{Homogeneous } In this strategy, we only consider homogeneous combinations of $\delta t$, i.e., combinations with just one value of $\delta t$. For example, for $T$ = 24 we consider [6, 6, 6, 6], [12, 12], and [24].

\textbf{Best \emph{m} in \emph{n} } We generate $n$ different, possibly heterogeneous combinations of $\delta t$,
validate each combination, and pick $m$ combinations with the lowest validation losses for testing.

The two strategies offer a trade-off between efficiency and expressivity. The homogeneous strategy only requires running three combinations for each lead time $T$, while best $m$ in $n$ provides greater expressivity.
Upon determining these combinations and executing the model rollouts, we obtain the final forecast by averaging the individual predictions. This approach achieves a similar effect to ensembling in NWP, where multiple forecasts are generated by running NWP models with different perturbed versions of the initial condition \citep{RootsofEnsembleForecasting}. As target lead times extend beyond 5--7 days and individual forecasts begin to diverge due to the chaotic nature of the atmosphere, averaging these forecasts is a Monte Carlo integration approach to handle this sensitivity to initial conditions and the uncertainty in the analyses used as initial conditions \citep{Metropolis}. We note that our inference strategy is distinguished from that used in Pangu-Weather: 
while Pangu-Weather trains a separate model for each time interval \(\delta t\), we train a single model for all \(\delta t\) values by conditioning on \(\delta t\). Additionally, while Pangu-Weather relies on a single combination of intervals to minimize rollout steps, our method improves forecast accuracy by averaging multiple forecasts derived from diverse combinations.

\subsection{Model architecture} \label{sec:arc}
We instantiate the framework in Section~\ref{sec:training} with a simple Transformer~\citep{vaswani2017attention}-based architecture.
Due to the similarity of weather forecasting to various dense prediction tasks in computer vision, one might consider applying Vision Transformer (ViT)~\citep{dosovitskiy2020image} for this task. However, weather data is distinct from natural images, primarily due to its significantly higher number of input channels, representing atmospheric variables with intricate physical relationships. For example, the wind fields are closely related to the gradient and shape of the geopotential field, and redistribute moisture and heat around the globe. Effectively modeling these interactions is critical to forecast accuracy. 
\subsection{Weather-specific embedding}
The standard patch embedding module in ViT, which uses a linear layer for embedding all input channels within a patch into a vector, may not sufficiently capture the complex interactions among input atmospheric variables. Therefore, we adopt for our architecture a weather-specific embedding module, consisting of two components, \emph{variable tokenization} and \emph{variable aggregation}.

\textbf{Variable tokenization } 
Given an input of shape $V \times H \times W$, variable tokenization linearly embeds each variable independently to a sequence of shape $(H / p) \times (W / p) \times D$, in which $p$ is the patch size and $D$ is the hidden dimension. We then concatenate the output of all variables, resulting in a sequence of shape $(H/p) \times (W/p) \times V \times D$.

\textbf{Variable aggregation }
We employ a single-layer cross-attention mechanism with a learnable query vector to aggregate information across variables. This module operates over the variable dimension on the output of the tokenization stage to produce a sequence of shape $(H/p) \times (W/p) \times D$. This module offers two primary advantages. First, it reduces the sequence length by a factor of $V$, significantly alleviating the computational cost as we use transformer to process the sequence. Second, unlike standard patch embedding, the cross-attention layer allows the models to learn non-linear relationships among input variables, enhancing the model's capacity to capture complex physical interactions. We present the complete implementation details of the weather-specific embedding in Section~\ref{sec:train_details}.

Figure~\ref{fig:compare_embeding} shows the superior performance of weather-specific embedding to standard patch embedding at all lead times from 1 to 10 days. A similar weather-specific embedding module was introduced by ClimaX~\citep{nguyen2023climax} to improve the model's flexibility when handling diverse data sources with heterogeneous input variables. We show that this specialized embedding module outperforms the standard patch embedding even when trained on a single dataset, due to its ability to model interactions between atmospheric variables through cross-attention effectively.

\subsubsection{\name{} Transformer block }
Following weather-specific embedding, the tokens are processed by a stack of transformer blocks~\citep{vaswani2017attention}. In addition to the input $X_0$, the block also needs to process the time interval $\delta t$. We do this by replacing the standard layer normalization module used in transformer blocks with adaptive layer normalization (adaLN)~\citep{perez2018film}. In adaLN, instead of learning the scale and shift parameters $\gamma$ and $\beta$ as independent parameters of the network, we regress them with an one-layer MLP from the embedding of $\delta t$. Compared to ClimaX~\citep{nguyen2023climax} which only adds the lead time embedding to the tokens before the first attention layer, adaLN is applied to every transformer block, thus amplifying the conditioning signal. Figure~\ref{fig:adaln} shows the consistent improvement of adaLN over the additive lead time embedding used in ClimaX. Adaptive layer norm was widely used in both GANs~\citep{karras2019style,brock2018large} and Diffusion~\citep{dhariwal2021diffusion,peebles2023scalable} to condition on additional inputs such as time steps or class labels. Figure~\ref{fig:arc} illustrates \name{}'s architecture. We refer to~\citep{nguyen2023climax} for illustrations of the weather-specific embedding block.

\section{Experiments}
\label{sec:exp}
\begin{figure*}[t!]
    \centering
    \includegraphics[width=0.9\textwidth]{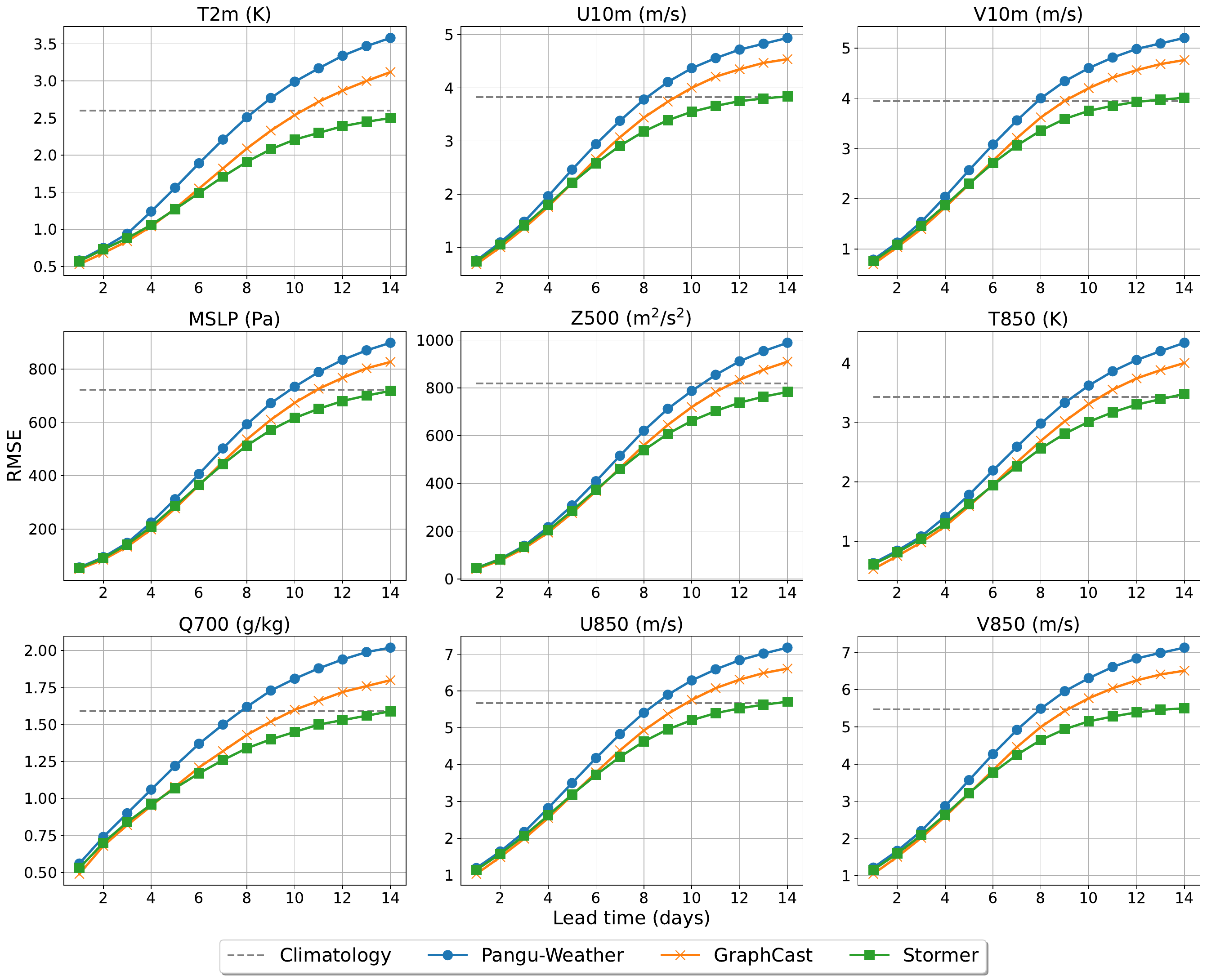}
    \caption{Global forecast results of \name{} and the baselines. We show the latitude-weighted RMSE for select variables. \name{} is on par or outperforms the baselines for the shown variables. During the later portion of the forecasts, \name{} gains $\sim1$ day of forecast skill with respect to climatology compared to the next best deep learning model. We note that \name{} was trained on much lower resolution data (1.40625$^\circ$) compared to Pangu-Weather (0.25$^\circ$) and GraphCast (0.25$^\circ$).
    }
    \label{fig:main_results}
\end{figure*}
We compare \name{} with state-of-the-art weather forecasting methods, and conduct extensive ablation analyses to understand the importance of each component in \name{}. We also study \name{} scalability by varying model size and the number of training tokens. We conduct all experiments on WeatherBench~2 (WB2)~\citep{rasp2023weatherbench}, a standard benchmark for data-driven weather forecasting. 

\textbf{Data:} We train and evaluate \name{} on the ERA5 dataset from WB2, which is the curated version of the ERA5 reanalysis data provided by ECMWF \citep{hersbach2020era5}. 
In its raw form, ERA5 contains hourly data from 1979 to the current time at 0.25$^\circ$ (721$\times$1440 grids) resolution, with different atmospheric variables spanning 137 pressure levels plus the Earth's surface. WB2 downsamples this data to 6-hourly with 13 pressure levels and provides different spatial resolutions. In this work, we use the 1.40625$^\circ$ (128$\times$256 grids) data. We use four surface-level variables -- 2-meter temperature (T2m), 10-meter U and V components of wind (U10 and V10), and Mean sea-level pressure (MSLP), and five atmospheric variables -- Geopotential (Z), Temperature (T), U and V components of wind (U and V), and Specific humidity (Q), each at $13$ pressure levels \{$50$, $100$, $150$, $200$, $250$, $300$, $400$, $500$, $600$, $700$, $850$, $925$, $1000$\}. We use 1979 to 2018 for training, 2019 for validation, and 2020 for testing.

\textbf{\name{} architecture:} For the main comparison in Section~\ref{sec:main_results}, we report the results of our largest \name{} model with 24 transformer blocks, 1024 hidden dimensions, and a patch size of 2, which is equivalent to ViT-L except for the smaller patch size. We vary the model size and patch size in the scaling analysis. For the remaining experiments, we report the performance of the same model as for the main result, but with a larger patch size of 4 for faster training.

\textbf{Training:} For the main result in Section~\ref{sec:main_results},
we train \name{} in three phases, as described in Section~\ref{sec:multi_step}. We train the model for 100 epochs for the first phase, 20 epochs for the second, and 20 epochs for the third. We perform early stopping on the validation loss aggregated across all variables, and evaluate the best checkpoint of the final phase on the test set. For the remaining experiments, we only train \name{} for the first phase due to computational constraints.

\textbf{Evaluation:} We evaluate \name{} and two deep learning baselines on forecasting nine key variables: T2m, U10, V10, MSLP, Z500, T850, Q700, U850, and V850. These variables are also used to report the headline scores in WB2. For each variable, we evaluate the forecast accuracy at lead times from 1 to 14 days, using the latitude-weighted root-mean-square error (RMSE) metric. 
For the main results, we use best $m$ in $n$ inference for rolling out \name{} as it yields the best result, with $m$ = 32 and $n$ = 128 chosen randomly from all possible combinations. For the remaining experiments, we use homogeneous inference for faster evaluations.
We provide results on the non-ensemble version of Stormer, probabilistic metrics with IC perturbations, a comparison between two inference strategies, and additional ablation studies in Appendix~\ref{sec:more_analysis}.

\subsection{Comparison with State-of-the-art models} \label{sec:main_results}
We compare the forecast performance of \name{} with Pangu-Weather~\citep{bi2023accurate} and GraphCast~\citep{lam2022graphcast}, two leading deep learning methods for weather forecasting. Pangu-Weather employs a 3D Earth-Specific Transformer architecture trained on the same variables as \name{}, but with hourly data and a higher spatial resolution of 0.25$^\circ$. GraphCast is a graph neural network that was trained on 6-hourly ERA5 data at 0.25$^\circ$, using 37 pressure levels for the atmospheric variables, and two additional variables, total precipitation and vertical wind speed. Both Pangu-Weather and GraphCast are iterative methods. GraphCast operates at 6-hour intervals, while Pangu-Weather uses four distinct models for 1-, 3-, 6-, and 24-hour intervals, and combines them to produce forecasts for specific lead times. We include Climatology as a simple baseline. We also compare \name{} with IFS HRES, the state-of-the-art numerical forecasting system, and IFS ENS (mean), which is the ensemble version of IFS. Since WB2 does not provide forecasts of these numerical models beyond $10$ days, we defer the comparison against these models to Appendix~\ref{sec:more_compare}. 

\textbf{Results:}
Figure~\ref{fig:main_results} evaluates different methods on forecasting nine key weather variables at lead times from $1$ to $14$ days.
For short-range, $1$--$5$ day forecasts, \name{}'s accuracy is on par with or exceeds that of Pangu-Weather, but lags slightly behind GraphCast.
\textit{At longer lead times, \name{} excels, consistently outperforming both Pangu-Weather and GraphCast from day 6 onwards by a large margin}. 
Moreover, the performance gap increases as we increase the lead time. 
At $14$ day forecasts, \name{} performs better than GraphCast by $10\%-20\%$ across all $9$ key variables.
\name{} is also the only model in this comparison that performs better than Climatology at long lead times, while other methods approach or even do worse than this simple baseline.
The model's superior performance at long lead times is attributed to the use of randomized dynamics training, which improves forecast accuracy by averaging out multiple forecasts, especially when individual forecasts begin to diverge.

Moreover, we also note that \name{} achieves this performance with much less compute and training data compared to the two deep learning baselines. We train \name{} on 6-hourly data of 1.40625$^\circ$ with 13 pressure levels, which is approximately 190$\times$ less data than Pangu-Weather's hourly data at 0.25$^\circ$ and 90$\times$ less than that used for GraphCast, which also uses 6-hourly data but at a 0.25$^\circ$ resolution with 37 pressure levels. The training of \name{} was completed in under 24 hours on 128 A100 GPUs. In contrast, Pangu-Weather took 60 days to train four models on 192 V100 GPUs, and GraphCast required 28 days on 32 TPUv4 devices. This training efficiency will facilitate future works that build upon our proposed framework. 

\begin{figure*}
  \centering
  \begin{subfigure}{0.35\linewidth}
    \includegraphics[width=1.0\linewidth]{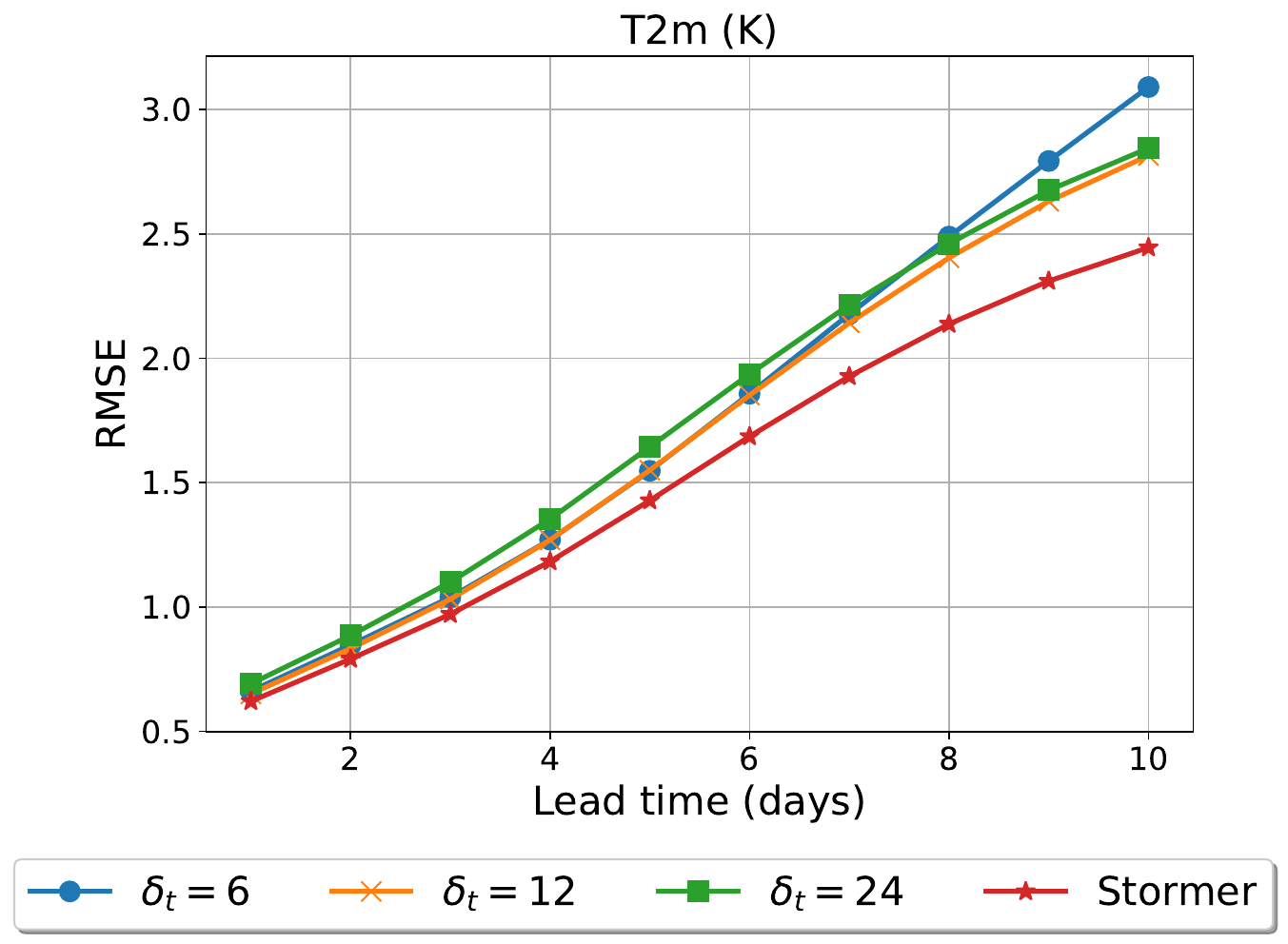}
    \caption{Impact of randomized forecasting.}
    \label{fig:randomized_effect}
  \end{subfigure}
  \hfill
  \begin{subfigure}{0.31\linewidth}
    \includegraphics[width=1.0\linewidth]{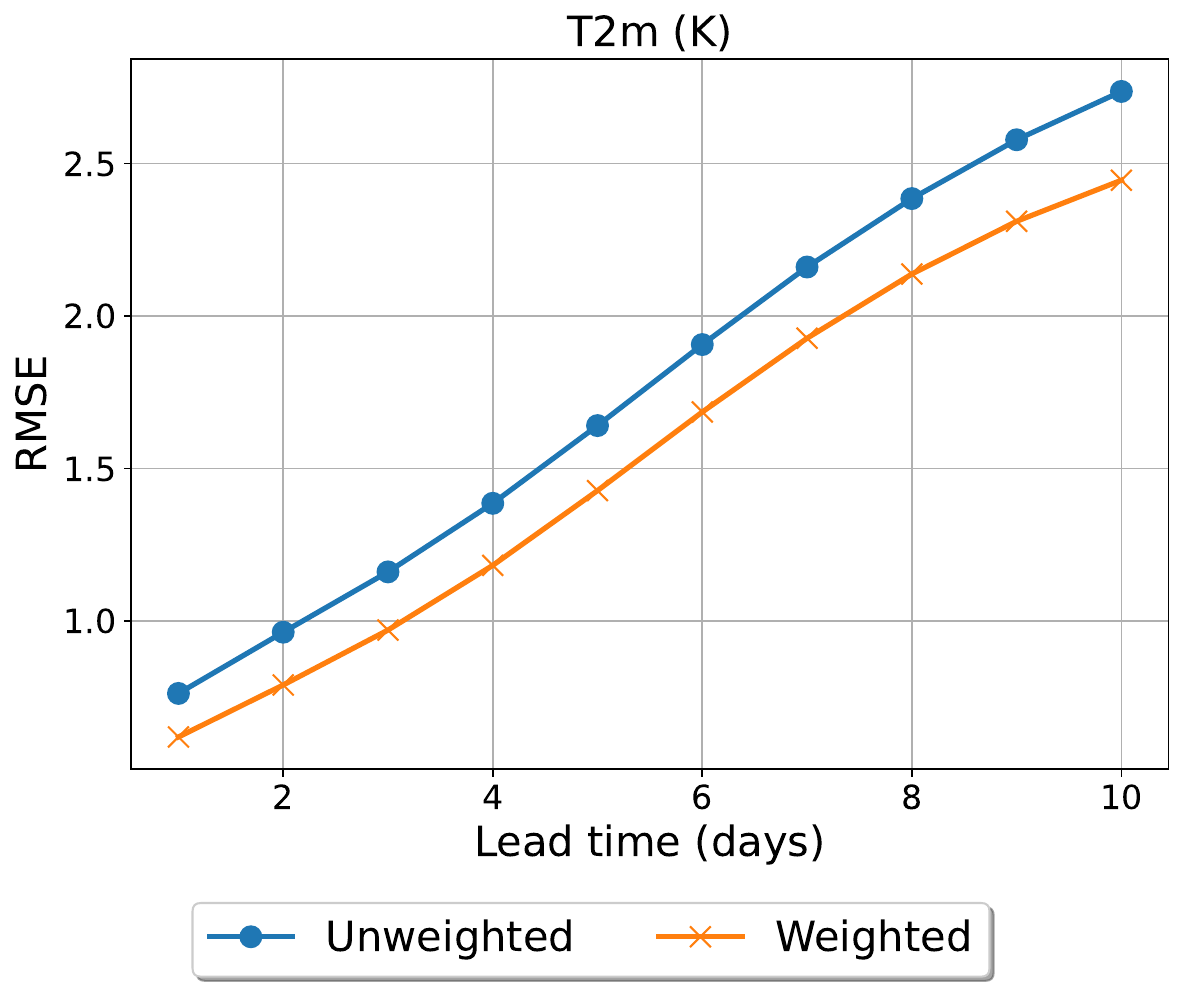}
    \caption{Impact of weighted loss.}
    \label{fig:weighting_effect}
  \end{subfigure}
  \hfill
  \begin{subfigure}{0.31\linewidth}
    \includegraphics[width=1.0\linewidth]{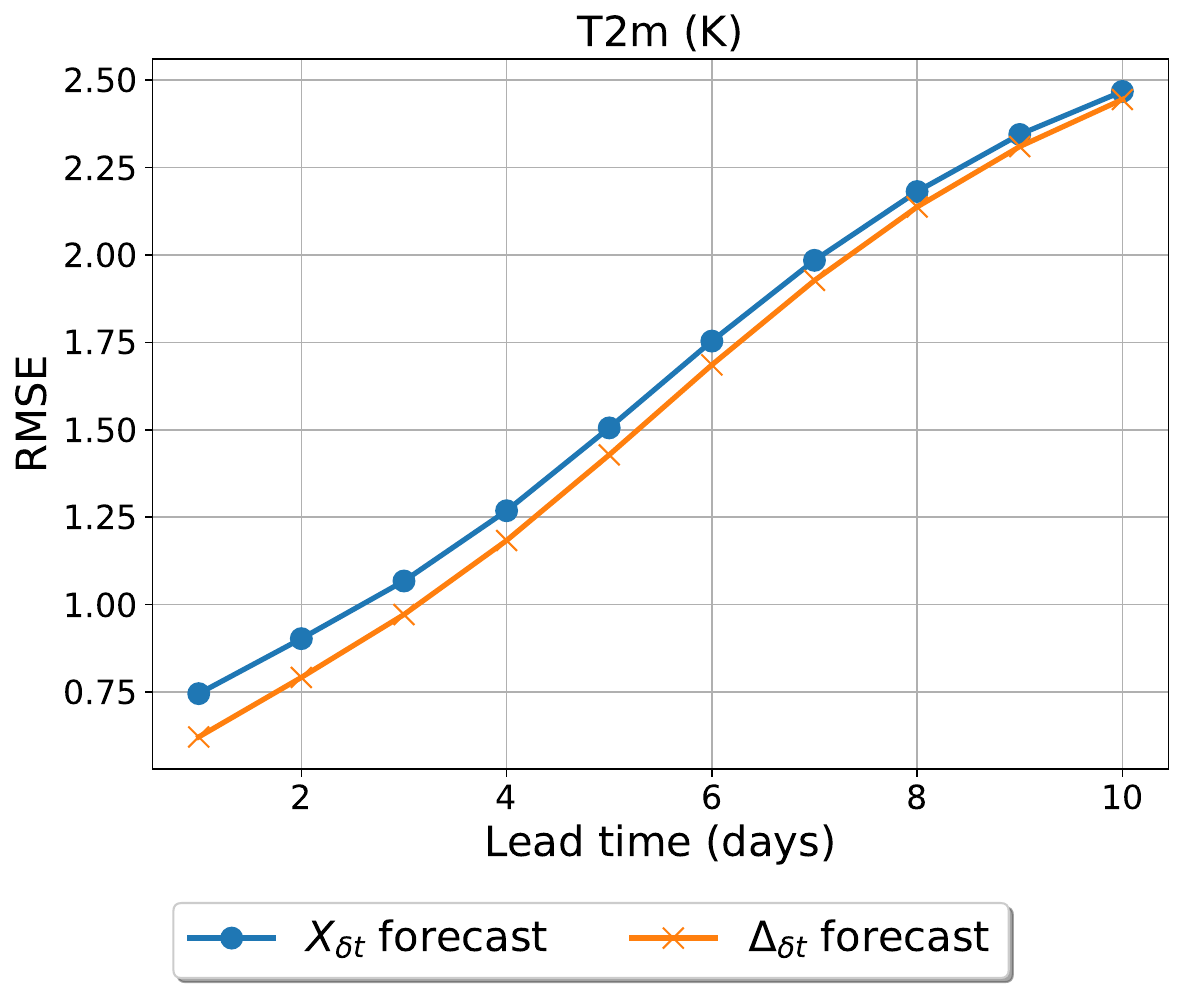}
    \caption{Absolute vs.\ dynamics forecast.}
    \label{fig:delta_vs_abs}
  \end{subfigure}
  \caption{Ablation studies showing the importance of different components in \name{}: (a) Randomized forecasting, (b) Pressure-weighted loss, and (c) Dynamics forecasting.}
  \label{fig:ablation_studies}
\end{figure*}

\subsection{Ablation studies} \label{sec:ablation}
We analyze the significance of individual elements within \name{} by systematically omitting one component at a time and observing the difference in performance.

\textbf{Impact of randomized forecasts:} 
We evaluate the effectiveness of our proposed randomized iterative forecasting approach. Figure~\ref{fig:randomized_effect} compares the forecast accuracy on surface temperature of \name{} and three
models trained with different values of $\delta t$. \name{} consistently outperforms all single-interval models at all lead times, and the performance gap widens as the lead time increases. We attribute this result to the ability of \name{} to produce multiple forecasts and combine them to improve accuracy. We note that \name{} achieves this improvement with no computational overhead compared to the single-interval models, as the different models share the same architecture and were trained for the same duration.

\textbf{Impact of pressure-weighted loss:}
Figure~\ref{fig:weighting_effect} shows the superior performance of \name{} when trained with the pressure-weighted loss. Intuitively, the weighting factor prioritizes variables that are nearer to the surface, as these variables are more important for weather forecasting and climate science.

\textbf{Dynamics vs.\ absolute forecasts:}
We justify our decision to forecast the dynamics $\Delta_{\delta t}$ by comparing with a counterpart that forecasts $X_{\delta t}$. Figure~\ref{fig:delta_vs_abs} shows that forecasting the changes in weather conditions (dynamics) is consistently more accurate than predicting complete weather states. One possible explanation for this result is that it is simpler for the model to predict the changes between two consecutive weather conditions than the entire state of the weather; thus, the model can focus on learning the most significant signal, enhancing forecast accuracy.

\begin{figure}[t]
  \centering
  \begin{subfigure}{0.32\linewidth}
    \includegraphics[width=1.0\linewidth]{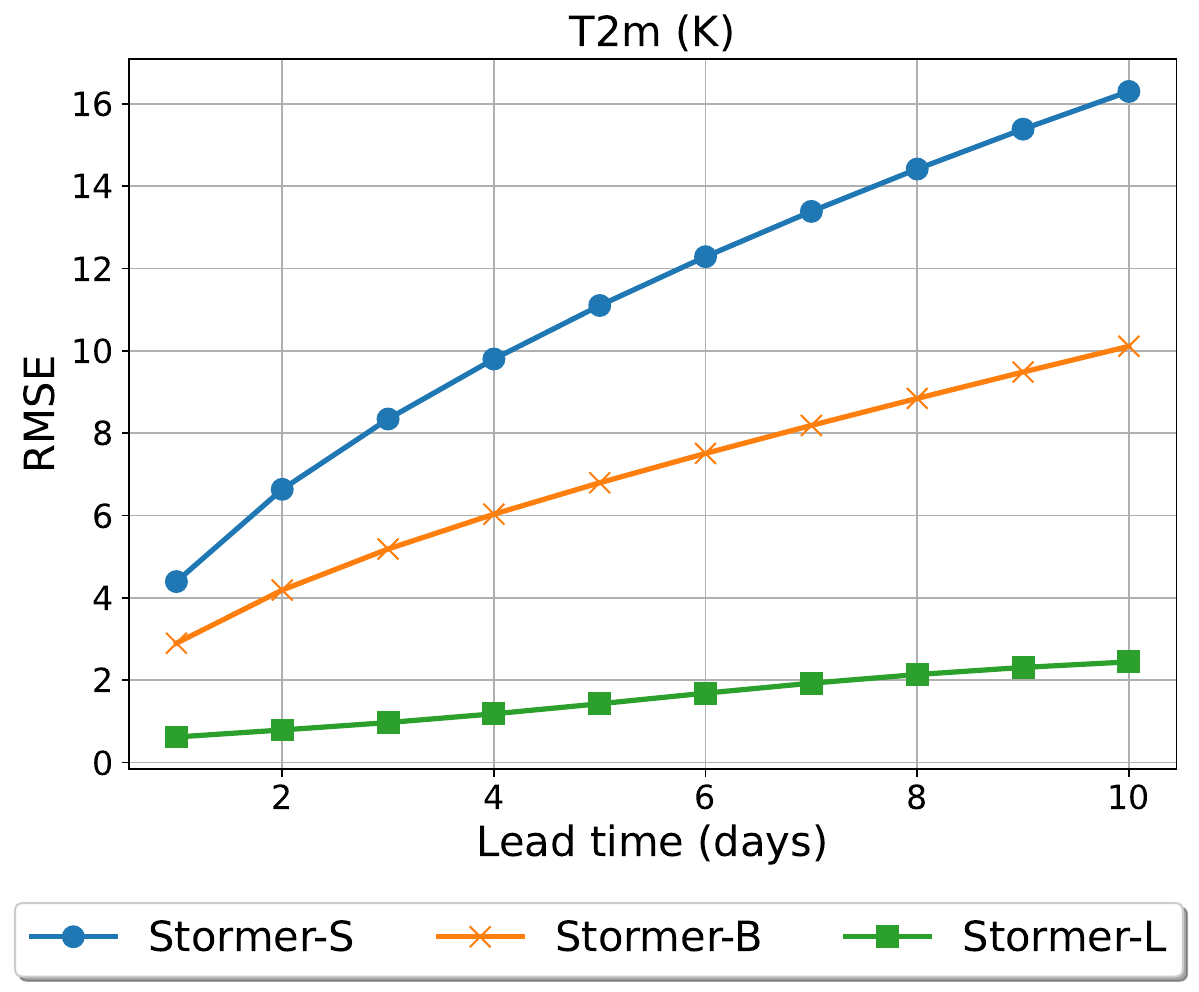}
  \end{subfigure}
  \hspace{0.3in}
  \begin{subfigure}{0.32\linewidth}
    \includegraphics[width=1.0\linewidth]{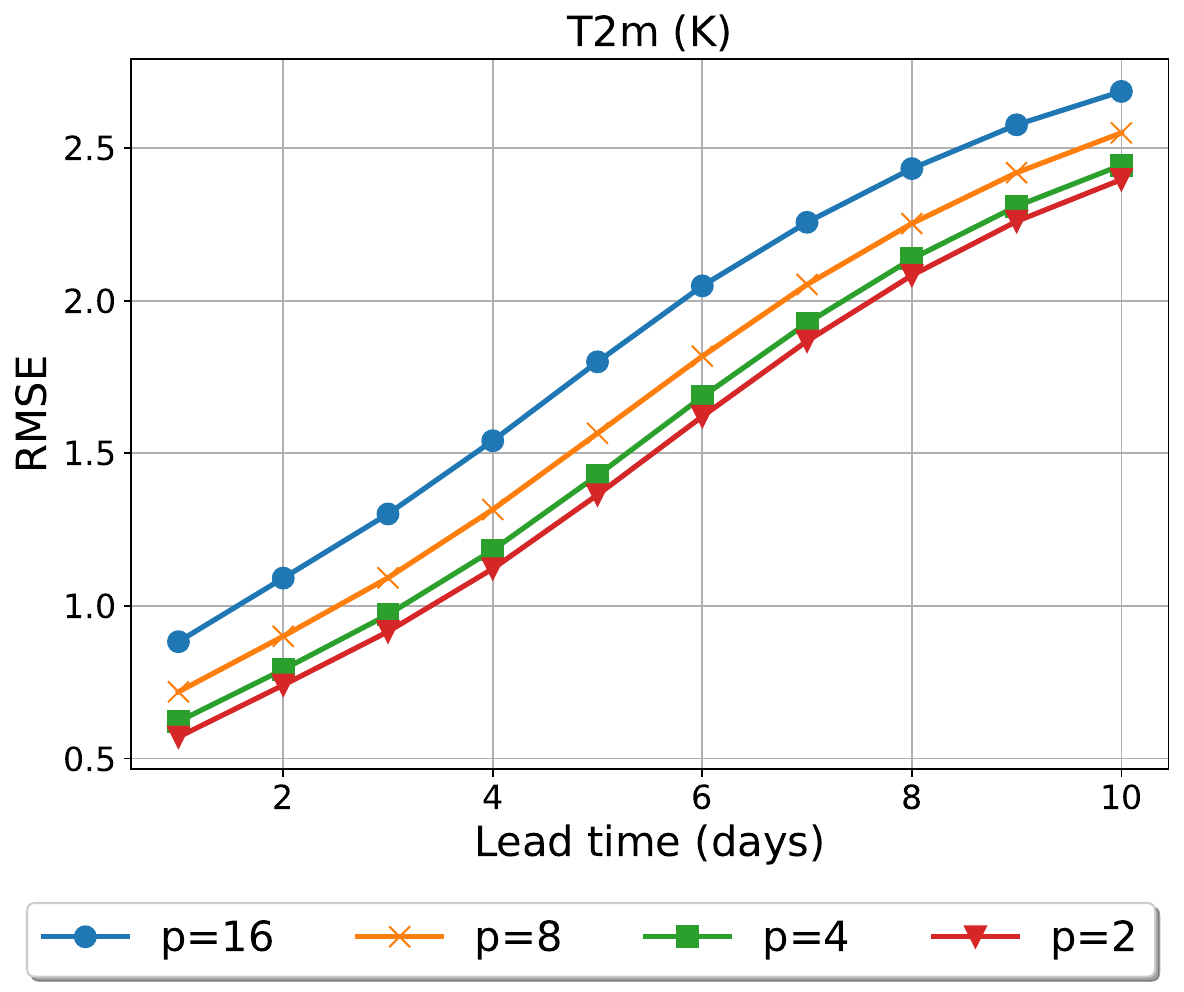}
  \end{subfigure}
  \caption{\name{} improves consistently with larger models (left) and smaller patch sizes (right).}
  \label{fig:scaling}
\end{figure}

\subsection{Scaling analysis}
We examine \name{}'s scalability in terms of model size and training tokens. We evaluate three variants -- \name{}-S, \name{}-B, and \name{}-L, with parameter counts similar to ViT-S, ViT-B, and ViT-L, respectively. To understand the impact of training token count, we vary the patch size from $2$ to $16$, quadrupling the training tokens each time the patch size is halved. Figure~\ref{fig:scaling} shows a significant improvement in forecast accuracy with larger models, and the performance gap widens with increased lead time. Since we do not perform multi-step fine-tuning for these models, minor performance differences at short intervals may magnify over time. While multi-step fine-tuning could potentially reduce this gap, it is unlikely to eliminate it entirely. Reducing the patch size also improves the performance of the model consistently. From a practical view, smaller patches mean more tokens and consequently more training data. From a climate perspective, smaller patches capture finer weather details and processes not evident in larger patches, allowing the model to more effectively capture physical dynamics that drive weather patterns.
\section{Related Work}
\label{sec:related_work}

\textbf{Deterministic weather forecasting }
Deep learning offers a promising approach to weather forecasting due to its fast inference and high expressivity. Early efforts~\citep{gmd-11-3999-2018,scher2018toward,weyn2019can} attempted training simple architectures on small weather datasets. To facilitate progress, WeatherBench~\citep{rasp2020weatherbench} provided standard datasets and benchmarks, leading to subsequent works that trained Resnet~\citep{he2016deep} and UNet architectures~\citep{weyn2020improving} for weather forecasting. 
These works showed the potential of deep learning but still displayed inferior accuracy to numerical systems.
However, significant improvements have been made in the last few years.~\citet{keisler2022forecasting} proposed a graph neural network (GNN) that performs iterative forecasting with $6$-hour intervals, performing comparably with some NWP models. FourCastNet~\citep{pathak2022fourcastnet} trained an adaptive Fourier neural operator and was the first neural network to run on 0.25$^\circ$ data. 
Pangu-Weather~\citep{bi2023accurate}, with its 3D Earth-Specific Transformer design, trained on high-resolution data, surpassed the benchmark IFS model. Following this, GraphCast~\citep{lam2022graphcast}  scaled up Keisler's GNN architecture to $0.25^\circ$, achieving even better results than Pangu-Weather. 
FuXi~\citep{Chen_fuxi_2023} was a subsequent work that trained a SwinV2~\citep{liu2022swin} on $0.25^\circ$ data and showed improvements over GraphCast at long lead times. However, FuXi requires finetuning multiple models specialized for different time ranges, increasing model complexity and computation. 
FengWu~\citep{chen2023fengwu} was a concurrent work with FuXi that also focused on improving long-horizon forecasts, but has not revealed complete model architecture and training details. ClimODE~\citep{vermaclimode} introduced physical inductive biases to provide better interpretability but was empirically inferior to existing methods.

\textbf{Probabilistic weather forecasting }
In addition to high accuracy, a desired ability of a weather forecasting model is to quantify forecast uncertainty. One common approach to achieve this is to combine an existing architecture with a probabilistic loss function. Gencast~\citep{price2023gencast} was one of the first works in this direction, combining the Graphcast architecture with a diffusion objective~\citep{ho2020denoising,karras2019style}, followed by Graph-EFM~\citep{oskarsson2024probabilistic}, which combined a hierarchical variant of Graphcast with the VAE objective~\citep{kingma2013auto}. This approach allows the model to generate multiple forecasts and estimate uncertainty after training. In an orthogonal approach, NeuralGCM~\citep{kochkov2024neural} proposed a hybrid forecasting system that combined a differentiable dynamical core with ML components for end-to-end training. The dynamical core allows the method to leverage powerful general circulation models and generate forecast ensembles via IC perturbations similar to NWP. However, the dynamical core in NeuralGCM is more computationally expensive than forward-passing a neural network and can limit the method's performance with an imperfect circulation model.

\section{Conclusion and Future Work}
\label{sec:conclusion}
This work proposes \name{}, a simple yet effective deep learning model for weather forecasting. We demonstrate that a standard vision architecture can achieve competitive results with a carefully designed training recipe. Our novel approach, randomized iterative forecasting, trains the model to forecast at different time intervals, enabling it to produce and combine multiple forecasts for each target lead time for better accuracy. Experiments show \name{}'s competitive accuracy in short-range forecasts and exceptional performance beyond 7 days, all with significantly less data and computing resources. Future research could explore using multiple forecasts to quantify uncertainty, randomizing other model components like input variables to increase variability and accuracy, and evaluating \name{} on higher-resolution data and larger model sizes due to its favorable scaling properties.

\section{Acknowledgments}
\label{sec:Acknowledgments}
AG acknowledges support from Google, Cisco, and Meta.
SM is supported by the U.S. Department of Energy, Office of Science, Advanced Scientific Computing Research, through the SciDAC-RAPIDS2 institute under Contract DE-AC02-06CH11357. RM and VK are supported under a Laboratory Directed Research and Development (LDRD) Program at Argonne National Laboratory, through U.S. Department of Energy (DOE) contract DE-AC02-06CH11357. TA is supported by the Global Change Fellowship in the Environmental Science Division at Argonne National Laboratory (grant no. LDRD 2023-0236). RM acknowledges support from DOE-FOA-2493: "Data intensive scientific machine learning". An award for computer time was provided by the U.S. Department of Energy’s (DOE) Innovative and Novel Computational Impact on Theory and Experiment (INCITE) Program and Argonne Leadership Computing Facility Director's discretionary award. This research used resources from the Argonne Leadership Computing Facility, a U.S. DOE Office of Science user facility at Argonne National Laboratory, which is supported by the Office of Science of the U.S. DOE under Contract No. DE-AC02-06CH11357.

\clearpage
\printbibliography

\appendix

\clearpage

\clearpage
\setcounter{page}{1}

\section{Borader impacts} \label{sec:broader}
Weather and climate modeling is crucial for understanding and tackling climate change. Creating better models based on deep learning could offer faster and cheaper alternatives to expensive numerical simulations. These models could improve weather predictions, extreme event forecasts, and climate projections. They might also help reduce the carbon footprint, better prepare for natural disasters, and enhance our knowledge of the Earth. However, relying only on deep learning models requires careful checks and monitoring, especially when predicting new or uncertain scenarios.

\section{Experiment details}
\label{sec:train_details}

\subsection{\name{} architecture}
\begin{figure}[h]
    \centering
    \includegraphics[width=0.6\linewidth]{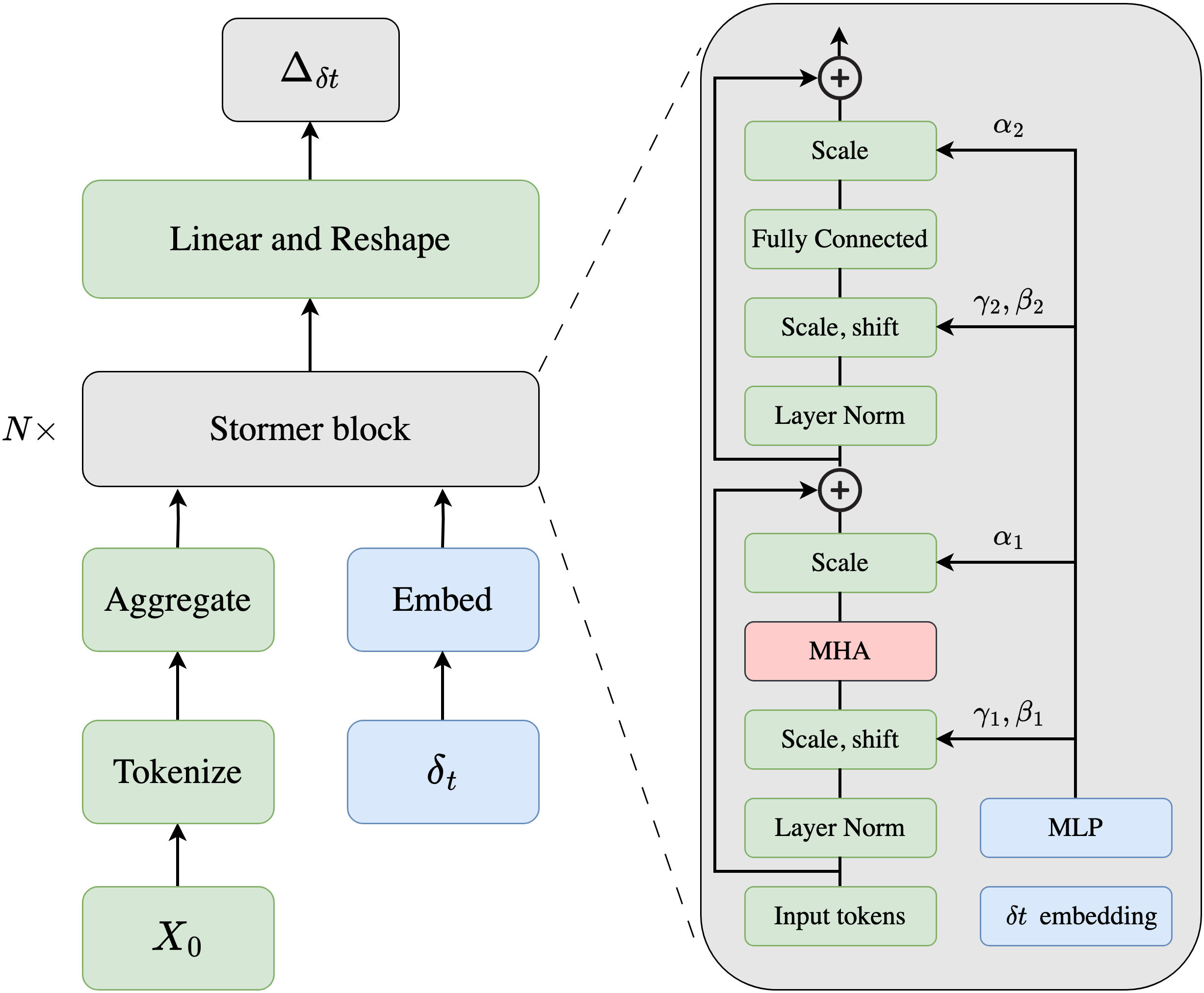}
    \caption{\name{} architecture. The initial condition goes through tokenization and aggregation, before being fed to a stack of $N$ \name{} blocks together with $\delta t$. Each \name{} block employs adaLN for $\delta t$ conditioning. 
    }
    \label{fig:arc}
\end{figure}

Figure~\ref{fig:arc} illustrates the architecture of \name{}. The variable tokenization module tokenizes each variable of the input $X_0 \in \mathbb{R}^{V\times H \times W}$ separately, resulting in a sequence of $V \times (H/p) \times (W/p)$ tokens, where $p$ is the patch size. The variable aggregation module then performs cross-attention over the variable dimension and outputs a sequence of $(H/p) \times (W/p)$ tokens. The interval $\delta t$ is embedded and fed to the \name{} backbone together with the tokens. The output of the last \name{} block is then passed through a linear layer and reshaped to produce the prediction $\Delta_{\delta t}$. Each \name{} block employs adaptive layer normalization to condition on additional information from $\delta t$. Specifically, the scale and shift parameters $(\gamma_1, \beta_1)$ and $(\gamma_2, \beta_2)$ are output by an MLP which takes $\delta t$ embedding as input. This MLP network additionally outputs $\alpha_1$ and $\alpha_2$ to scale the output of the attention and fully connected layers, respectively.

In all experiments, the variable tokenization module is a standard patch embedding layer usually used in ViT, and the aggregation module is a single-layer multi-head cross-attention. The first embedding of $\delta t$ is a linear layer, and the adaLN module in each block employs a 2-layer MLP.
For the main comparison with the current methods, we train a \name{} model with a patch size of $2$, $1024$ hidden dimensions, and $24$ \name{} blocks. For the scaling experiments, we vary the hidden dimensions, number of blocks, and patch size. For the rest of the ablation studies, we use a patch size of $4$, hidden dimension of $1024$, and $24$ blocks.

\subsection{Training and evaluation details}

\subsubsection{Data normalization}
\paragraph{Input normalization } 
We compute the mean and standard deviation for each variable in the input across all spatial positions and all data points in the training set. This means each variable is associated with a scalar mean and scalar standard deviation. During training, we standardize each variable by subtracting it from the associated mean and dividing it by the standard deviation.

\paragraph{Output normalization } 
Unlike the input, the output that the model learns to predict is the difference between two consecutive steps. Therefore, for each variable, we compute the mean and standard deviation of the difference between two consecutive steps in the training set. What it means to be "consecutive" depends on the time interval $\delta t$. If $\delta t = 6$, we collect all pairs in training data that are 6-hour apart, compute the difference between two data points in each pair, and then compute the mean and standard deviation of these differences. Since we train \name{} with randomized $\delta t$, we repeat the same process for each value of $\delta t$.

\subsubsection{Three-phase training}
As mentioned in Section~\ref{sec:method}, we train \name{} in three phases with the following objective:
\begin{equation}
    \mathcal{L}(\theta) = \mathbb{E}\left[ \frac{1}{KVHW} \sum_{k=1}^K \sum_{v=1}^V \sum_{i=1}^H \sum_{j=1}^W w(v) L(i) (\widehat{\Delta}_{k \delta t}^{vij} - \Delta_{k \delta t}^{vij})^2 \right],
\end{equation}
where the number of rollout steps $K$ is equal to $1$, $4$, and $8$ in phase $1$, $2$, and $3$, respectively. For phases $2$ and $3$, we finetune the best checkpoint from the preceding phase. 

\subsubsection{Pressure weights}
For pressure-level variables, we assign weights proportionally to the pressure level of each variable. For $4$ surface variables, we assign $w = 1$ for T2m and $w = 0.1$ for the remaining variables U10, V10, and MSLP. The surface weights were proposed by GraphCast~\citep{lam2022graphcast} and we did not perform any additional hyperparameter tuning.

\subsubsection{Optimization}
For the $1$st phase, we train the model for $100$ epochs. We optimize the model using AdamW~\citep{kingma2014adam} with learning rate of $5e-4$, parameters $(\beta_11 = 0.9, \beta_2 = 0.95)$ and weight decay of $1e-5$. We used a linear warmup schedule for $10$ epochs, followed by a cosine schedule for $90$ epochs.

For the $2$nd and $3$rd phases, we train the model for $20$ epochs with a learning rate of $5e-6$ and $5e-7$, respectively. We used a linear warmup schedule for $5$ epochs, followed by a cosine schedule for $15$ epochs. Other hyperparameters remain the same.

We perform early stopping for all phases, where the criterion is the validation loss aggregated across all variables at lead times of $1$ day, $3$ days, and $5$ days for phases $1$, $2$, and $3$, respectively. We save the best checkpoint for each phase using the same criterion.

\subsubsection{Software and hardware stack}
We use PyTorch~\citep{paszke2019pytorch}, Pytorch Lightning~\citep{falcon2019pytorch}, timm~\citep{rw2019timm}, numpy~\citep{harris2020array} and xarray~\citep{Hoyer_2017} for data processing and model training. We trained \name{} on $128$ $40$GB A$100$ devices. We leverage mixed-precision training, Fully Sharded Data Parallel, and gradient checkpointing to reduce memory.

\subsection{Evaluation protocol}
As different models are trained on different resolutions of data, we follow the practice in WB2 to regrid the forecasts of all models to the same resolution of $1.40625^\circ$ ($128 \times 256$ grid points). We then calculate evaluation metrics on this shared resolution. Similarly to WB2, we evaluate forecasts with initial conditions at $00/12$UTC for all days in 2020.

\begin{figure*}[t]
    \centering
    \includegraphics[width=0.95\textwidth]{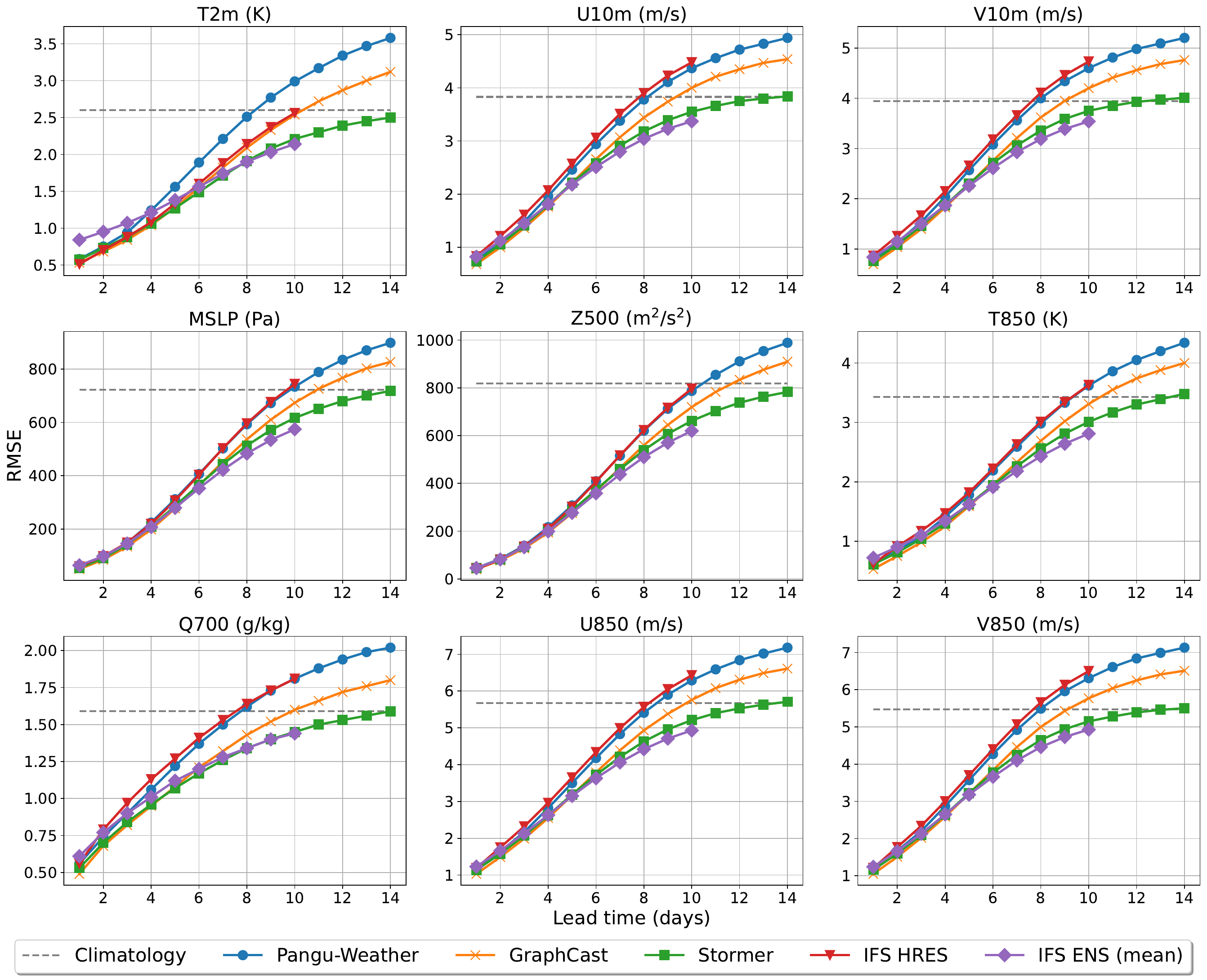}
    \caption{Global forecast verification results of \name{} and the baselines from $1$- to $14$-day lead times. We show the latitude-weighted RMSE for select variables. \name{} is on par or outperforms each of the benchmark models for the shown variables. During the later portion of the forecasts, \name{} significantly outperforms the current methods.
    }
    \label{fig:rmse_full}
\end{figure*}

\section{Additional results}
\label{sec:more_analysis}

\subsection{Complete comparison with SoTA models} \label{sec:more_compare}

Figure~\ref{fig:rmse_full} compares \name{} with both deep learning and numerical methods. We take IFS and IFS ENS from WB2 which is only available until day $10$. Similar to its deep learning counterparts, \name{} achieves lower RMSE compared to the IFS model for most variables, except for near-surface temperature (T2m) at initial lead times, and only performs slightly worse than IFS ENS. To the best of our knowledge, \name{} is the first model trained on 1.40625$^\circ$ data to surpass IFS.

\begin{figure*}[t]
    \centering
    \includegraphics[width=0.95\textwidth]{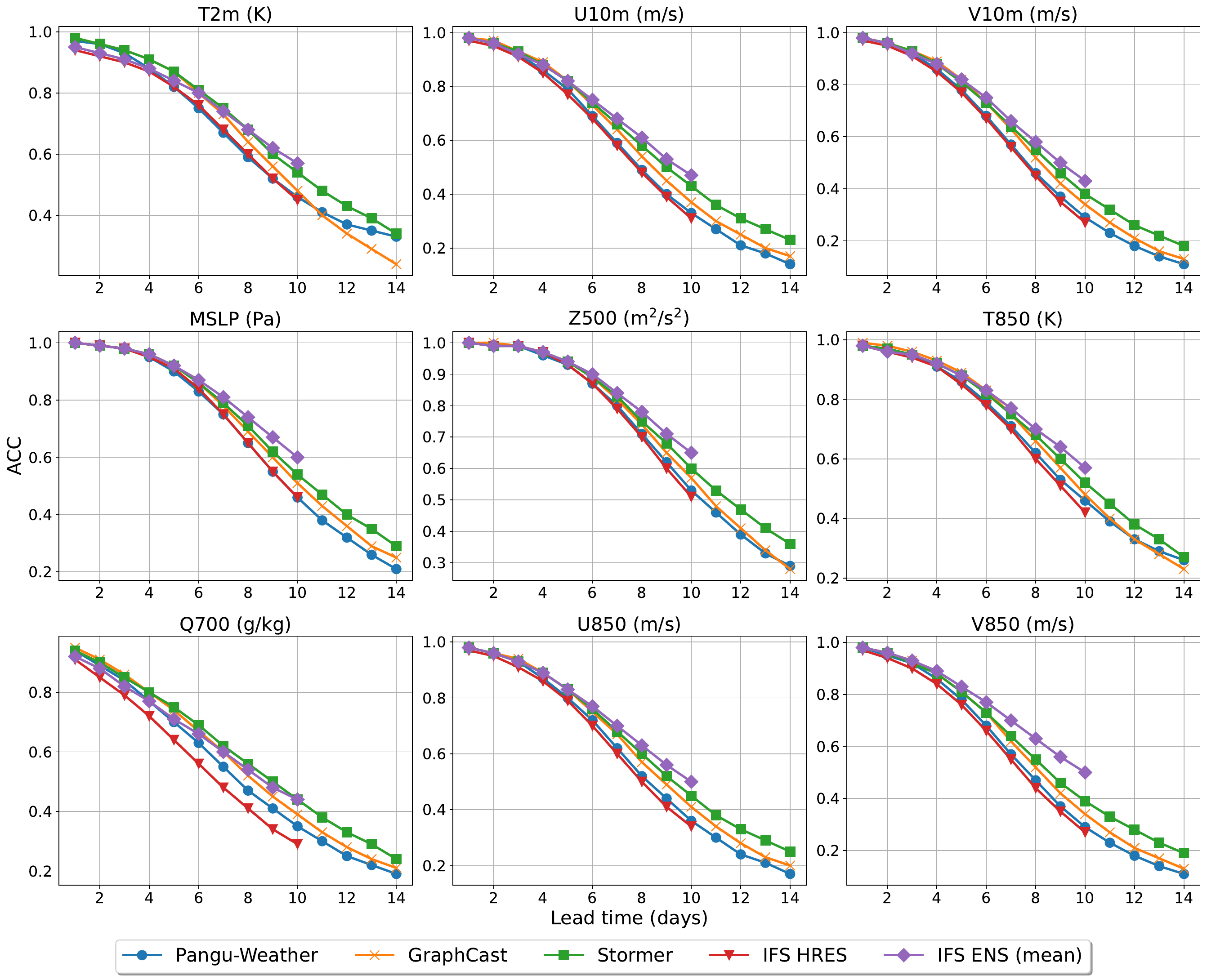}
    \caption{Global forecast verification results of \name{} and the baselines from $1$- to $14$-day lead times. We show the latitude-weighted ACC for select variables. \name{} is on par or outperforms each of the benchmark models for the shown variables. During the later portion of the forecasts, \name{} significantly outperforms the current methods.
    }
    \label{fig:acc_full}
\end{figure*}

Additionally, we compare \name{} and the baselines on latitude-weighted ACC, another common verification metric for weather forecast models. ACC represents the Pearson correlation coefficient between forecast anomalies relative to climatology and ground truth anomalies relative to climatology. ACC ranges from $-1$ to $1$, where $1$ indicates perfect correlation, and $-1$ indicates perfect anti-correlation. We refer to WB2~\citep{rasp2023weatherbench} for the formulation of ACC. Figure~\ref{fig:acc_full} shows that similarly to RMSE, \name{} achieves competitive performance from $1$ to $5$ days, and outperforms the baselines by a large margin beyond $6$ days.

\subsection{Impact of multi-step fine-tuning}
\begin{figure*}[h]
    \centering
    \includegraphics[width=0.7\textwidth]{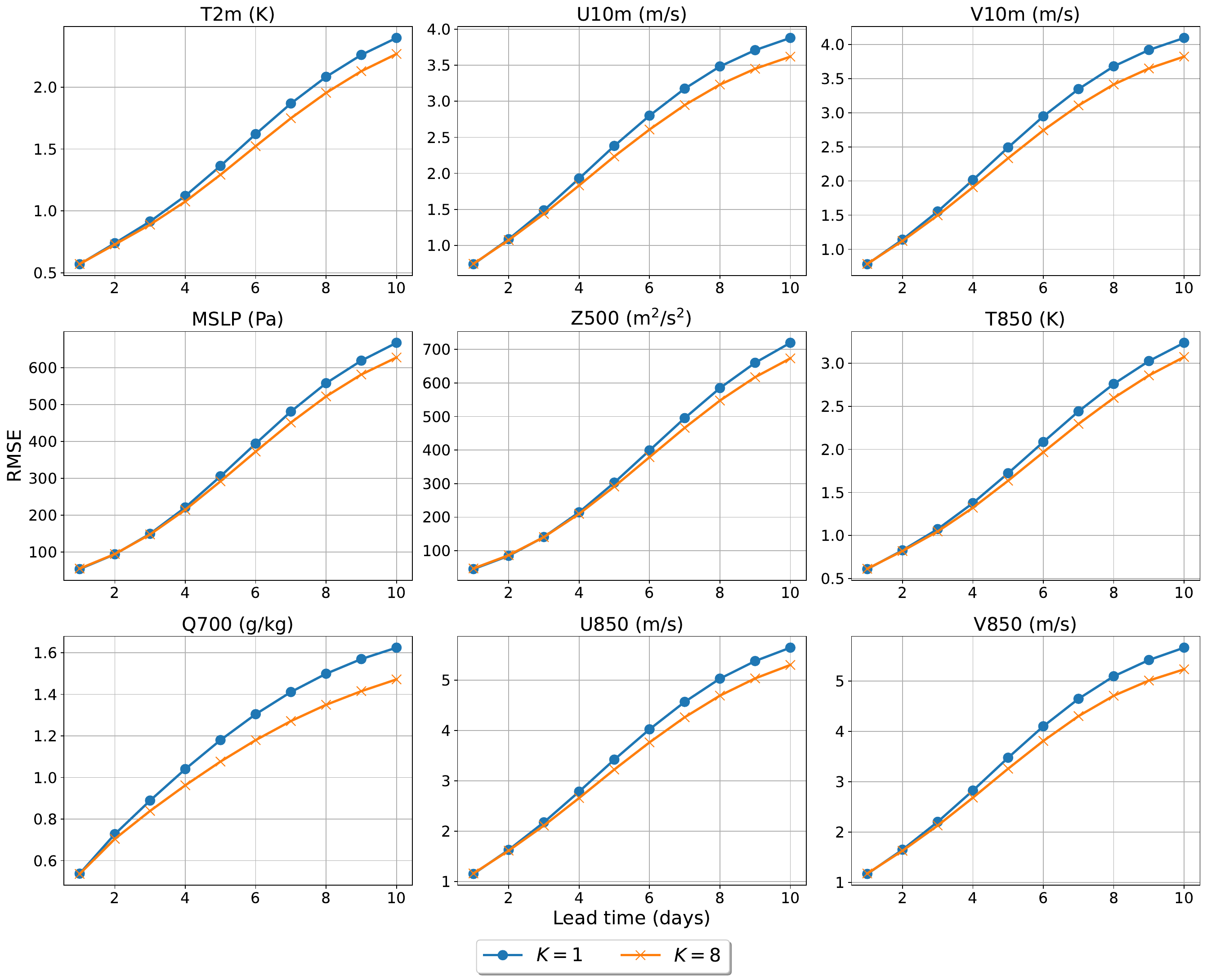}
    \caption{Performance of \name{} without ($K=1$) and with ($K=8$) multi-step fine-tuning.
    }
    \label{fig:multi_step_finetune}
\end{figure*}

We verify the importance of multi-step fine-tuning by comparing \name{} after the $1$st phase ($K=1$) and after the $3$rd phase ($K=8$). Figure~\ref{fig:multi_step_finetune} shows that multi-step fine-tuning significantly improves performance at long lead times.
\begin{figure*}[h!]
    \centering
    \includegraphics[width=0.85\textwidth]{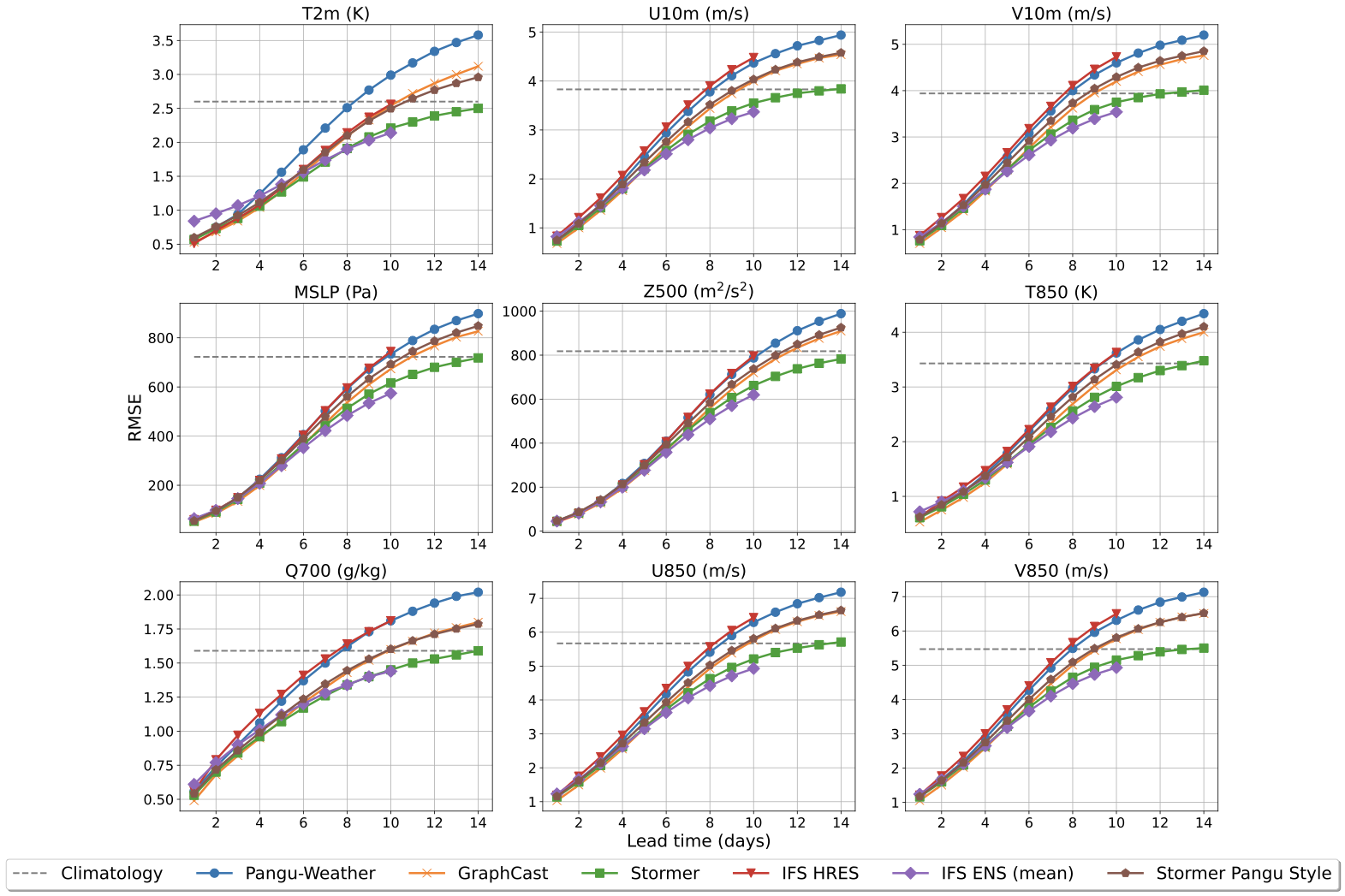}
    \caption{Non-ensemble version Stormer vs the baselines. }
    \label{fig:non_ensemble}
\end{figure*}

\subsection{Non-ensemble performance of Stormer}
Even though our inference strategy can be considered ensemble forecasting, we note that it is much cheaper and more efficient than common techniques such as training multiple networks, dropout, or IC perturbations, as we only have to train a single neural network and do not need extensive hyperparameter tuning. However, to provide more insights into the performance of Stormer, we additionally compare the non-ensemble version of Stormer with the baselines. Specifically, we performed the Pangu-style inference, where we only used the 24-hour interval forecasts to roll out into the future, instead of combining different intervals. Figure~\ref{fig:non_ensemble} shows that non-ensemble Stormer outperforms Pangu and performs competitively with Graphcast.

\subsection{Probabilistic forecasting with IC perturbations}

Since Stormer can produce forecast ensembles after training, we can consider it a probabilistic forecast system. However, our preliminary results suggested that different forecasts from Stormer are underdispersive and should not be used for uncertainty estimation. To make Stormer a probabilistic forecast system, we need to introduce more randomization to the forecasts via IC perturbations. To do this, for each combination of intervals during the Best m in n inference, we added 4 different noises sampled from a Gaussian distribution, resulting in a total of 128 ensemble members.

\begin{figure*}[h]
    \centering
    \includegraphics[width=0.9\textwidth]{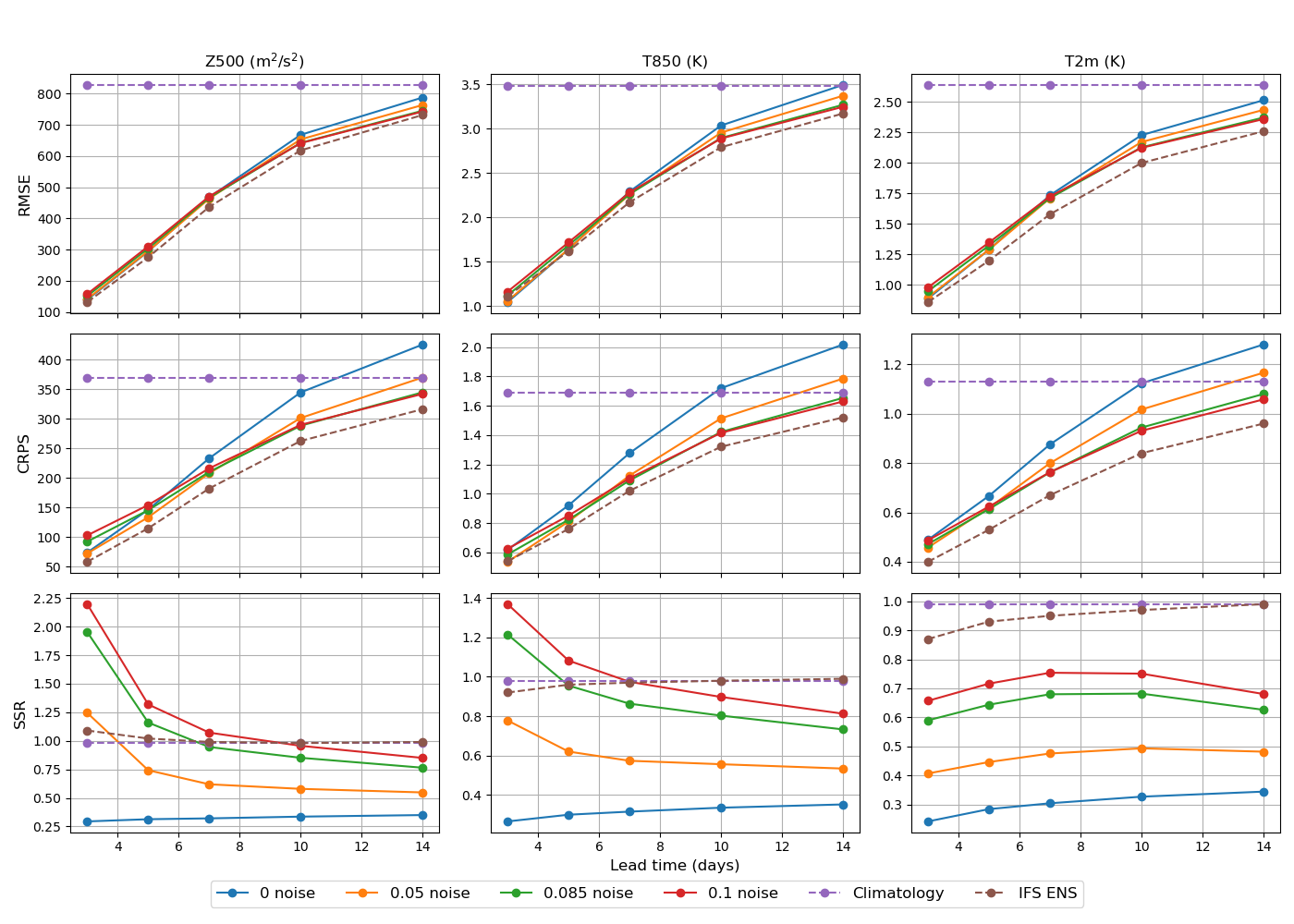}
    \caption{Probabilistic performance of Stormer with different levels of IC perturbations.}
    \label{fig:prob}
\end{figure*}

Figure~\ref{fig:prob} shows that IC perturbations improve the probabilistic metric significantly, but may hurt the deterministic performance at short lead times. Moreover, it is difficult to find an optimal noise level for the spread-skill ratio across different variables and lead times. We can further improve this by using a better noise distribution or variable-dependent and lead-time-dependent noise scheduling, which we defer to future works. 

\subsection{Comparison of different inference strategies}
\begin{figure*}[h]
    \centering
    \includegraphics[width=0.9\textwidth]{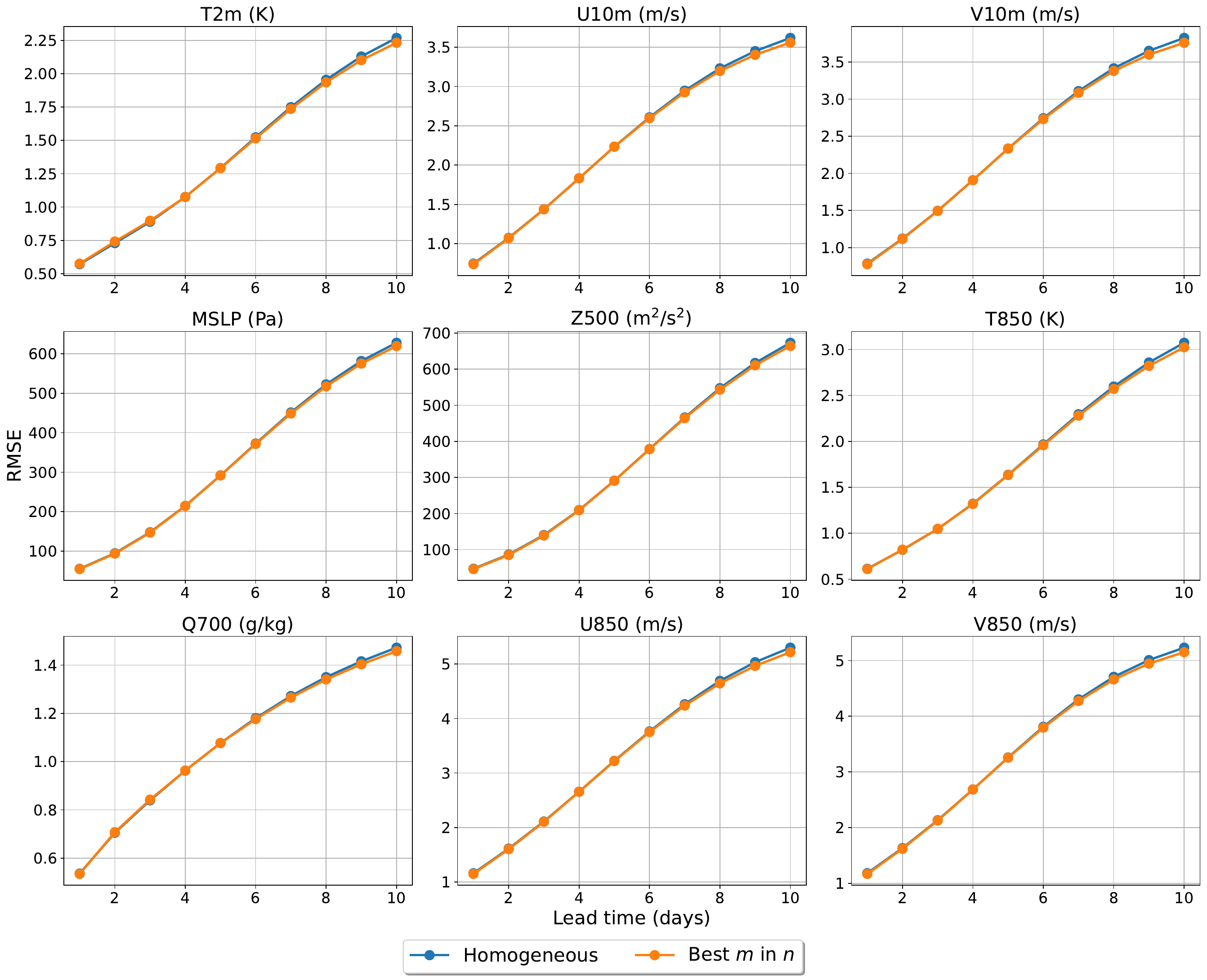}
    \caption{Comparsion of Homogeneous vs Best $m$ in $n$ inference strategies.}
    \label{fig:inference_compare}
\end{figure*}

Our two inference strategies, Homogeneous and Best $m$ in $n$, provide a tradeoff between efficiency and forecast accuracy. Figure~\ref{fig:inference_compare} compares the performance of these two strategies across different variables at different lead times. The results show that Homogeneous performs competitively with Best $m$ in $n$, while being much more efficient, requiring only 3 forward passes compared to $n$.

\clearpage
\subsection{Qualitative results}
We visualize forecasts produced by \name{} at lead times from $1$ days to $14$ days for $9$ key variables. All forecasts are initialized at $0$UTC January $26$th 2020. Each figure illustrates one lead time, where each row is for each variable. The first column shows the initial condition, the second column shows the ground truth at that lead time, the third column shows the forecast, and the last column shows the bias, which is the difference between the forecast and the ground truth.

\begin{figure*}[h]
    \centering
    \includegraphics[width=1.0\textwidth]{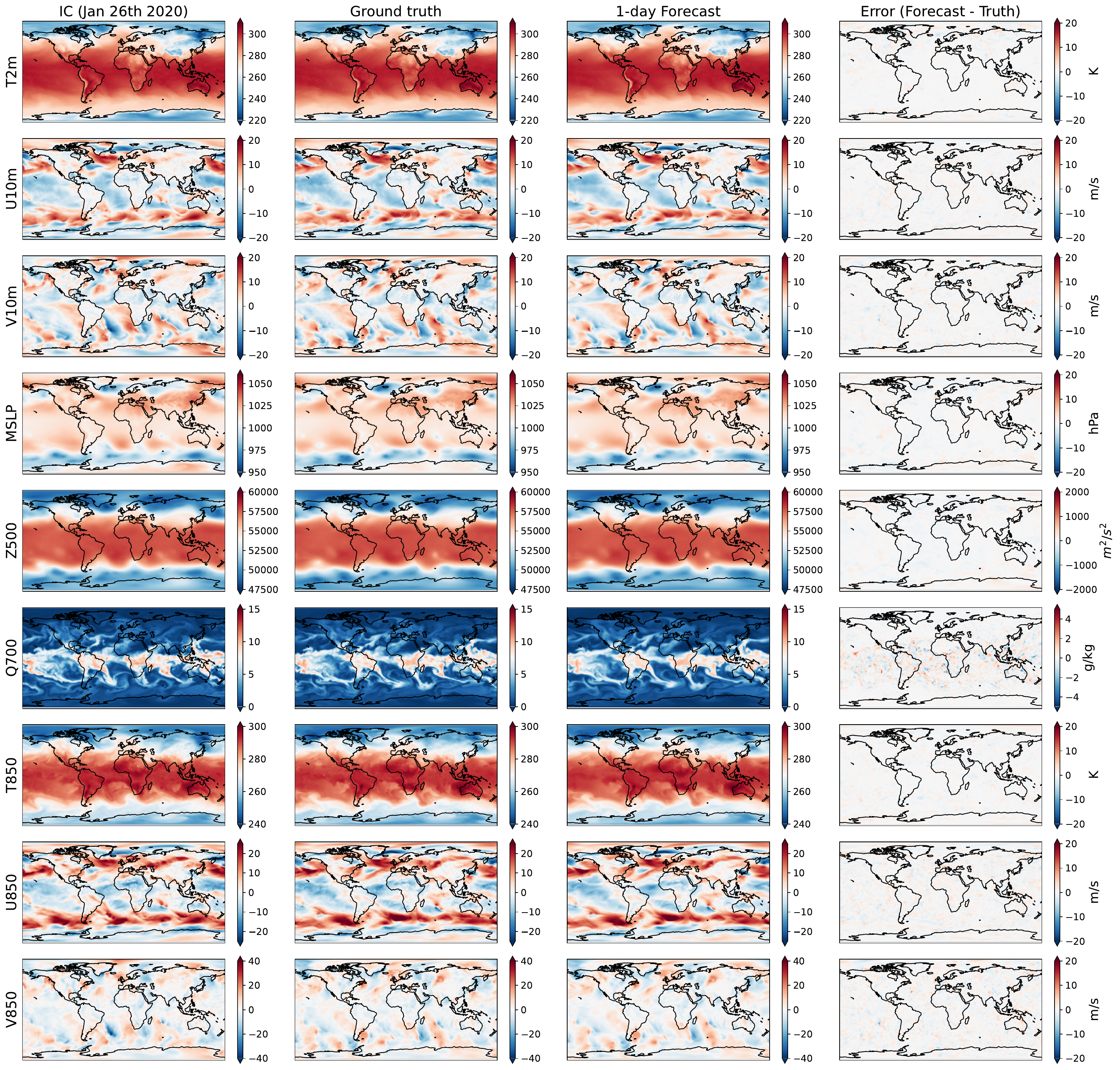}
    \caption{1-day lead time}
\end{figure*}

\begin{figure*}[t]
    \centering
    \includegraphics[width=1.0\textwidth]{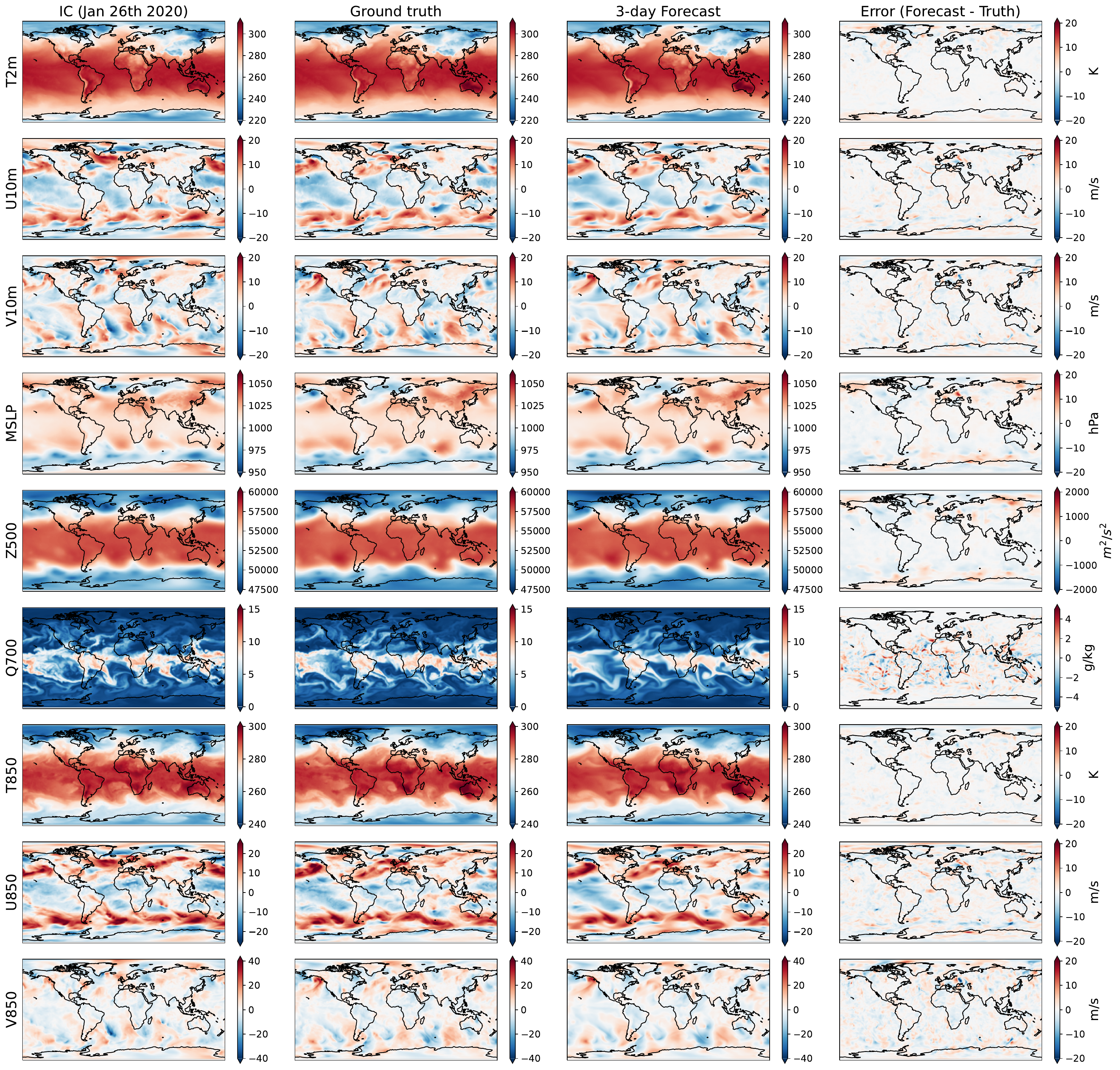}
    \caption{3-day lead time}
\end{figure*}

\begin{figure*}[t]
    \centering
    \includegraphics[width=1.0\textwidth]{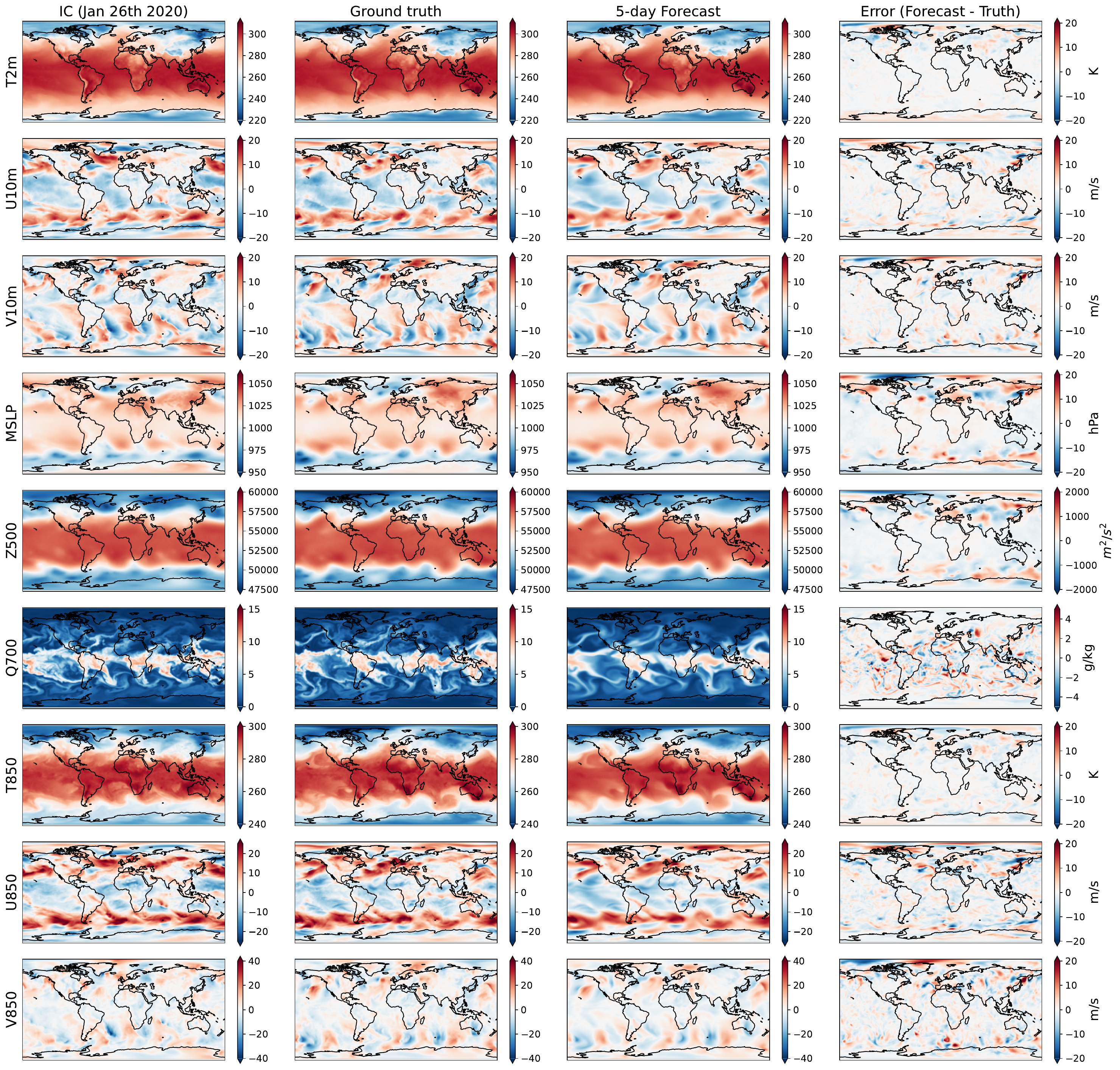}
    \caption{5-day lead time}
\end{figure*}

\begin{figure*}[t]
    \centering
    \includegraphics[width=1.0\textwidth]{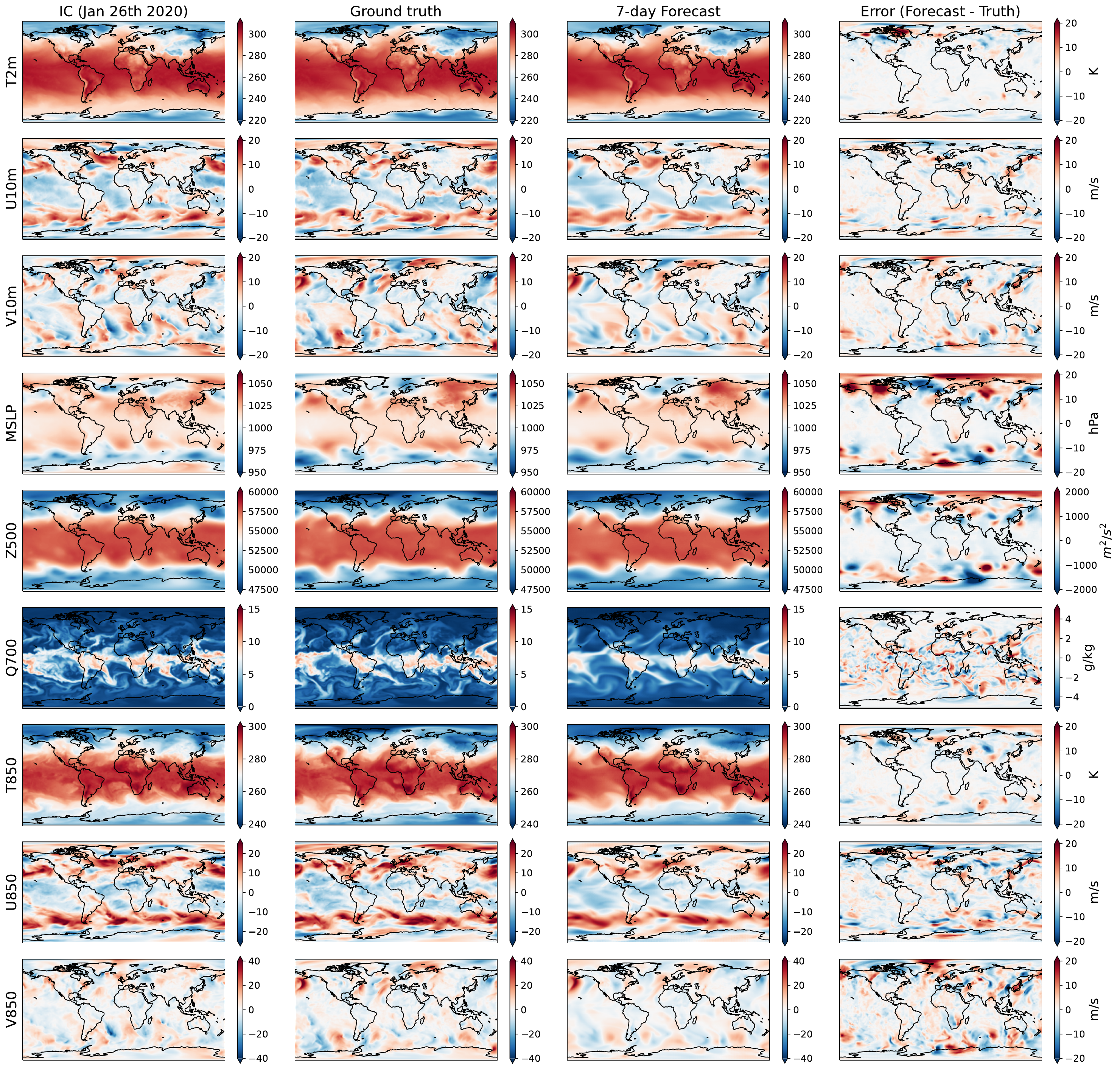}
    \caption{7-day lead time}
\end{figure*}

\begin{figure*}[t]
    \centering
    \includegraphics[width=1.0\textwidth]{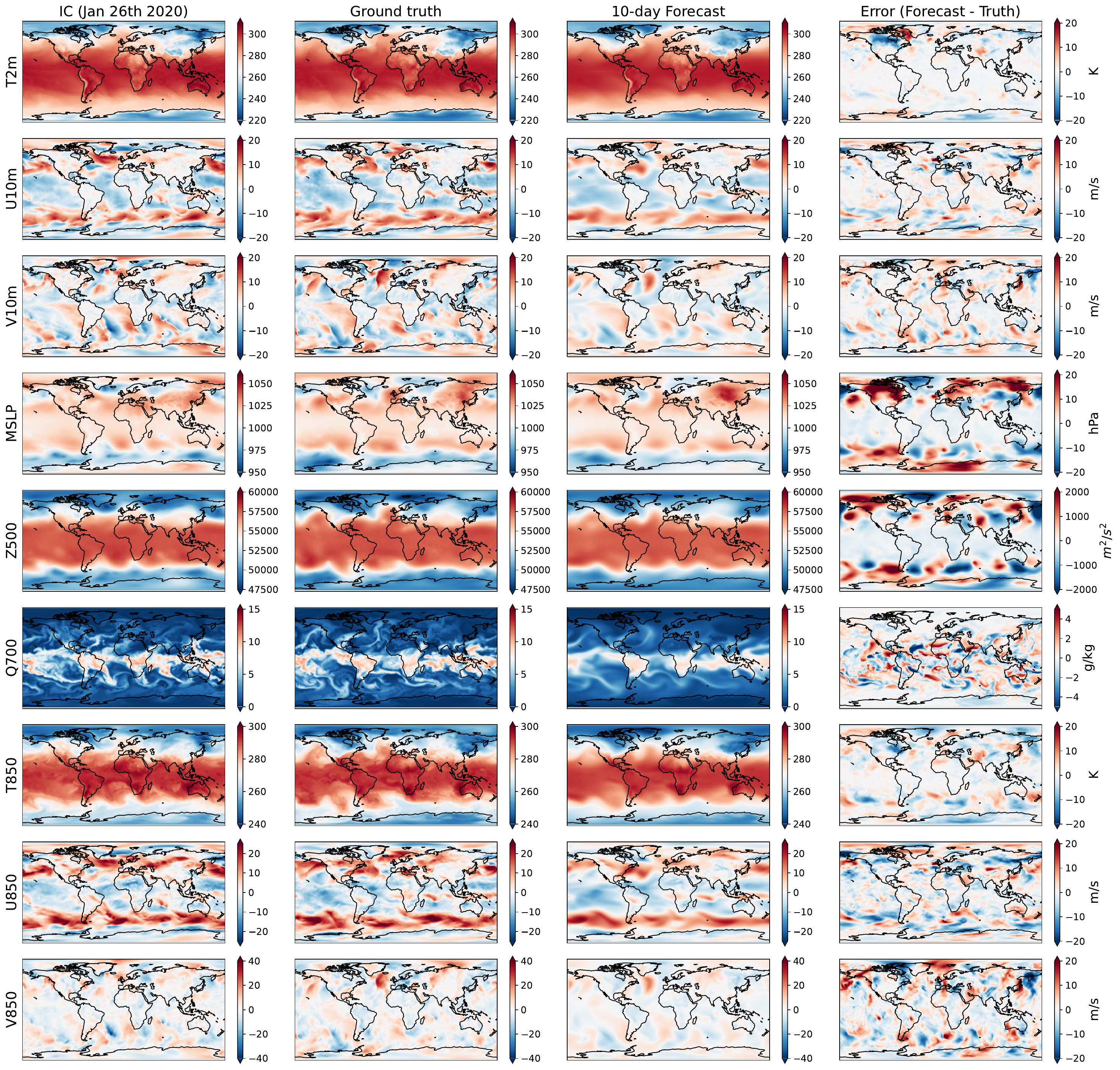}
    \caption{10-day lead time}
\end{figure*}

\begin{figure*}[t]
    \centering
    \includegraphics[width=1.0\textwidth]{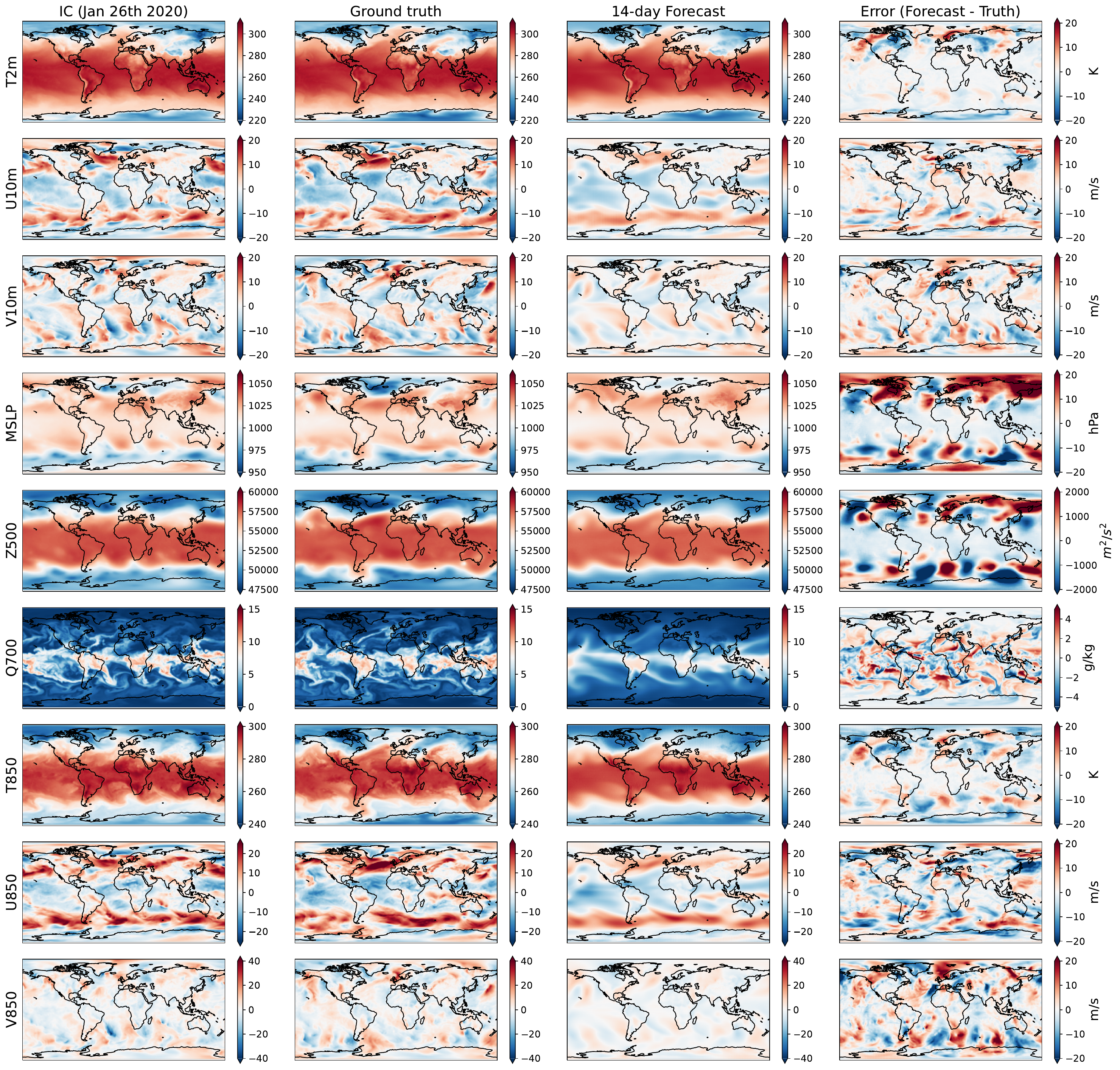}
    \caption{14-day lead time}
\end{figure*}

\end{document}